# Metal-Polypyridyl Complexes in Electronic Circuits

Rajwinder Kaur[a], Bijai Singh,[a] Vikram Singh,*,[b] and Prakash Chandra Mondal*,[a]

[a]Department of Chemistry, Indian Institute of Technology Kanpur, Uttar Pradesh-208016, India

[b]Department of Chemistry, Korea Advanced Institute of Science and Technology, Daejeon-34141, South Korea

**Abstract**: The integration of functional molecules onto conductive or dielectric surfaces represents a promising avenue for employing molecule-centric technologies, encompassing sensor development, electrochromism, and the facilitation of charge and spin transport at the nanoscale. These assemblies exhibit robust physical and chemical stability, exerting influence not only on the interfacial properties of pertinent surfaces but also on the characteristics of the assembled molecules themselves. Within the spectrum of available molecules, metal-polypyridyl complexes have emerged as focal points of investigation for generating surface-bound assemblies, encompassing monolayers or multilayers, achieved through pre- or post-functionalization methodologies. The appeal of these complexes lies in their inherent versatility, stemming from facile synthesis, appreciable stability, and tuneable properties encompassing redox behaviours, reactivity in excited states, luminescent emissions, excited state longevity, as well as responsiveness to stimuli-induced electron and energy transfer. This comprehensive review aims to delve into the nuanced traits and applications of surface-bound metal-polypyridyl complexes, elucidating their characteristics and practical implementations through pertinent and illustrative examples in the field of molecular electronics.

# 1. Introduction

The investigation of a molecule-surface interactions stands as a pivotal domain within contemporary surface science, captivating attention, and scrutiny.[1–3] These interactions encompass a spectrum of covalent and non-covalent associations contingent upon the intricate chemistry governing the union between functional molecules and the substrate.[4,5] Such unions yield either self-assembled layers or covalently assembled layers driven by self-assembly, each imbued with distinctive characteristics. The scope of molecular engineering encompasses a diverse array of entities, spanning from purely organic to inorganic or hybrid structures such as coordination and organometallic compounds.[6] Substrates, either metallic or semiconducting, wield substantial influence, affording deliberate control over the resultant ensembles properties. Notably compelling is the potential for the creation of meticulously ordered molecular arrays boasting densities nearing $10^{13}$ molecules per $cm^2$, materializing as either monolayers or extending into multilayer configurations spanning the nanometer scale. At this diminutive scale, these assemblies represent an extensive toolkit, offering insight into the fundamental facets of molecule-surface interaction and interface engineering. Moreover, their significance lies in the prospect of sculpting molecular-based electronics, spintronics, electrochromism, heterogeneous catalysis, and sensorics, thereby fostering groundbreaking opportunities within these domains.[7–9]

Throughout history, pivotal strides in the development of monolayers were traced back to Langmuir's seminal exploration into the behavior of amphiphiles on water surfaces, pioneering the conceptual framework for one-molecule-thick layers, each measuring approximately $10^{-7}$ cm, which could be meticulously organized into ordered monolayers.[10] Subsequently, Blodgett's groundbreaking discovery illuminated the possibility of transferring these layers from the air/water interface to solid substrates, leading to the inception of Langmuir-Blodgett (LB) films, albeit their initial characterization as thermodynamically unstable.[11] However, it was the work of Bigelow et al. that marked a transformative milestone by unveiling stable and spontaneous self-assembled monolayers (SAMs), achieved through the successful immobilization of alkylamines onto a platinum substrate.[12] The driving impetus behind the formation of these SAMs lay in the robust interactions between the adsorbate and the substrate, engendering ordered and densely packed structures through weak interactions among the adsorbed molecules. Extensive investigations have delved into an assortment of substrates and molecules to construct SAMs, revealing a common thread among adsorbate molecules characterized by the presence of an anchoring group (e.g., sulfur on gold substrate, silanes on $SiO_2$-based substrate, carboxylates on metal oxide substrate), a spacer group typically comprising a series of methylene units, and a terminal group typically representing the functional moiety (**Figure 1**).[13–15] While comprehending the fundamental underpinnings of SAMs remains pivotal, the domain of their applications has grown rapidly.[16,17] The interactions between the adsorbate molecule and substrate, as well as intermolecular interactions, bestow essential chemical and thermal stability, precise orientation, and avenues for ion/electron-transfer channels.[18,19] Yet, it is the functional component that endows these assemblies with the capacity for specific activities and functionalities.

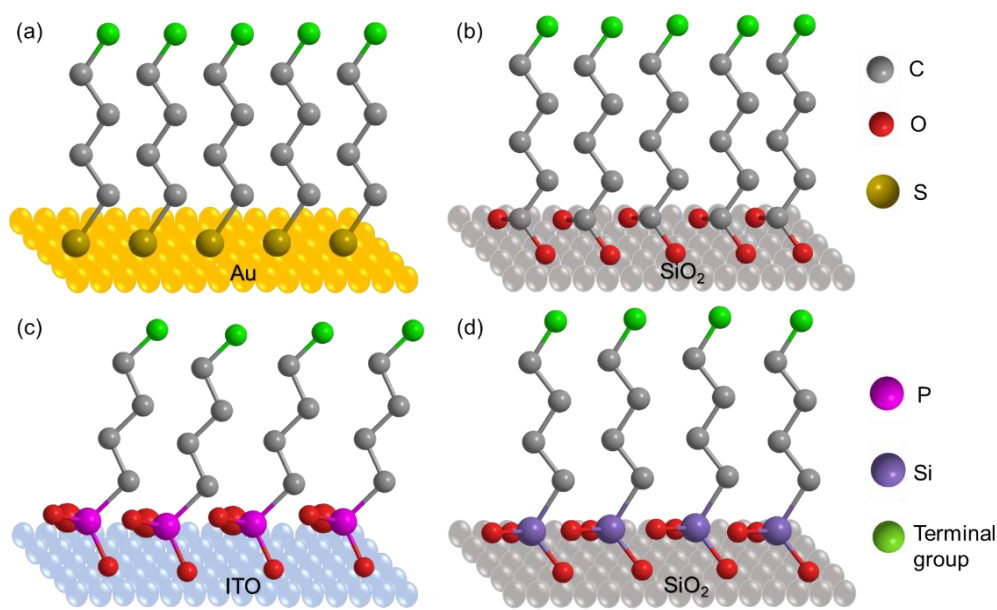

**Figure** 1. Schematics for representative self-assembled monolayers of molecules bearing different anchoring groups on different substrates; a) alkane thiols on gold, b) carboxylates on Si/SiO$_2$, c) phosphonic acid on ITO and, d) silanes on Si/SiO$_2$.

The scope of designing functional molecules is vast and intricate, yet metal-ligand coordination assemblies stand out as exceptionally versatile constructs. Within this paradigm, a plethora of metal ions, each offering distinct coordination tendencies dictated by oxidation states, can be paired with ligands possessing variable denticity. This affords the ability to craft meticulously pre-designed assemblies onto desired substrates. In this context, the surface chemistry of polypyridyl complexes has permeated across multidisciplinary domains such as synthetic and applied chemistry, surface engineering, nanotechnology, and molecular electronics. These complexes are composed of primary and secondary building blocks—the former represented by a metal ion characterized by specific size, oxidation state, and coordination number, and the latter encompassing polypyridyl ligands whose geometric arrangement, structural configuration, bond energies, and number of coordinating sites can be tailored. The judicious combination of a suitable metal ion and designer polypyridyl ligands endows these assemblies with augmented or supplementary physicochemical attributes. When these molecules are confined to well-defined substrates in monolayer or oligomer architectures via ether covalent or non-covalent interactions, a remarkable yet adjustable transformation of their properties ensues. This alteration arises from intimate interactions between these molecules and the substrate, fostering a diverse array of options in molecular design, surface-attachment chemistry, and resultant monolayer/oligomer characteristics. These extensive possibilities have forged novel pathways for applications in diverse fields such as molecular engineering, nanotechnology, sensing modules, electrochromism, non-volatile memory devices, and solid-state devices, among others (**Figure 2**). Herein, we will review the charge transport properties of metal-polypyridyl complexes through relevant literature examples involving molecular electronics-based charge-transport, photocurrent, and molecular memory developments.

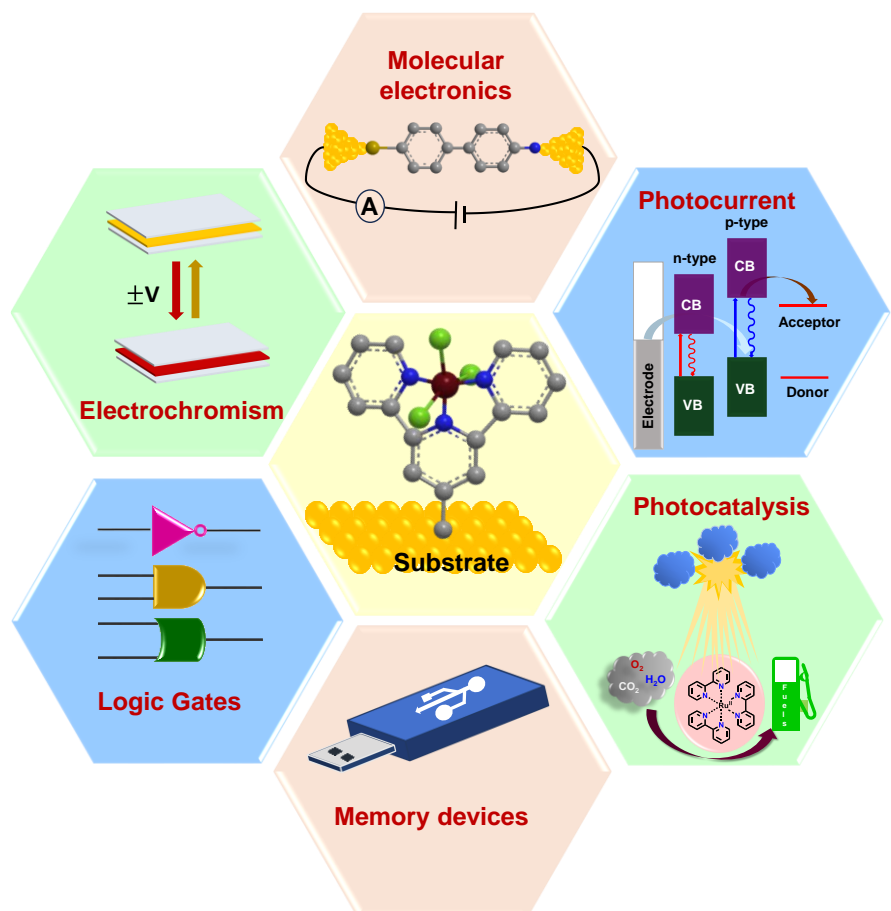

**Figure 2**: Potential applications of surface-confined metal-polypyridyl complexes.

## 2. Molecular Electronics: Charge Transport

The core of molecular electronics resides in molecular circuits.[20–22] These circuits, integral to any molecular device, rely on molecular wires supported by conducting solid surfaces for their connectivity. To effectively regulate molecular devices, it becomes imperative to synthesize, manage, and comprehend the mechanisms of charge transport within these molecular wires. A spectrum of methodologies exists for the formation of solid-state molecular junctions in such systems, encompassing traditional approaches like chemical vapor deposition or sputtering, conductive atomic force microscopy, nanopore junctions, and the mechanical control break junction (MCBJ), among other techniques.[23–26] Once the junction is established, the sample is primed for the exploration of transport mechanisms. The movement of charge within these wires is contingent upon the Fermi levels of the electrode and the highest occupied molecular orbital (HOMO) or lowest unoccupied molecular orbital (LUMO) of the molecules tethered to them. However, within transition metal-based molecular wires, two distinct modes of charge transfer have been observed: tunnelling and hopping, both extensively documented in the existing literature. In the case of tunneling (or super-exchange), the gap between the Fermi level of the electrodes and the conduction orbitals of the molecules is considerable, resulting in electrons swiftly traversing through the molecular orbitals with minimal residence time on the molecule. This tunneling current displays an exponential decrease concerning the distance between the electrodes, as characteristic of tunneling phenomena. Mathematically, tunneling current (*I*) in molecular wires varies according to the equation given below (**equation 1**):

$$I = I_O \, e^{-\beta d} \ldots\ldots(\text{i})$$

where *d* represents the length of the molecular wire and *β* denotes the tunneling decay constant, extensive documentation exists regarding the behavior of organic wires. It has been observed that tunneling occurs in organic wires when their size measures less than 3 nm. However, a transition to distinct charge transport phenomena—termed hopping—is witnessed within the range of 3 to 4 nm, as established by numerous studies. Furthermore, the level of conjugation also plays a crucial role in this transition; heightened conjugation within organic structures tends to accentuate the hopping characteristics in these wires. Notably, molecular wires involving transition metal complexes or metal center wires exhibit disparities compared to organic wires. In cases where similar organic ancillaries are present, the inclusion of metal centers augments the current flowing through these wires. Additionally, tunneling phenomena are infrequently observed within MCWs, marking a significant divergence from the behavior typically observed in organic wire systems. Incoherent hopping occurs in MCWs and the conductance (*G*) follows the Arrhenius type of relation:

$$G \propto \exp\left(\frac{-E_A}{k_B T}\right) \ldots\ldots(\text{ii})$$

Here, $E_A$ denotes the hopping activation energy, signifying that hopping necessitates thermal activation for its occurrence. The proximity between the Fermi levels of electrodes and the conduction orbitals (HOMO or LUMO) is narrow in this scenario. Consequently, charge transport takes place in discrete steps, akin to traversing stairs, moving from the electrodes to the conduction orbitals associated with different metal centers, and eventually reaching the other electrode (**Figure 3**). This sequential movement reflects a stepwise transport mechanism resembling the progression from one level to another on a staircase, facilitating the charge transfer from one terminal to the other. Within metal center molecular wires, two distinct types of hopping mechanisms have been observed (**Figure 4**): intramolecular directional hopping and non-directional diffusion hopping. Existing literature elucidates that in the case of intramolecular directional hopping, the current (*I*) varies inversely with the number of active redox sites ($N$) ($I \propto N^{-1}$) present on the MCWs. Conversely, for non-directional diffusion hopping, the current varies inversely with the square of the active redox sites ($I \propto N^{-2}$).

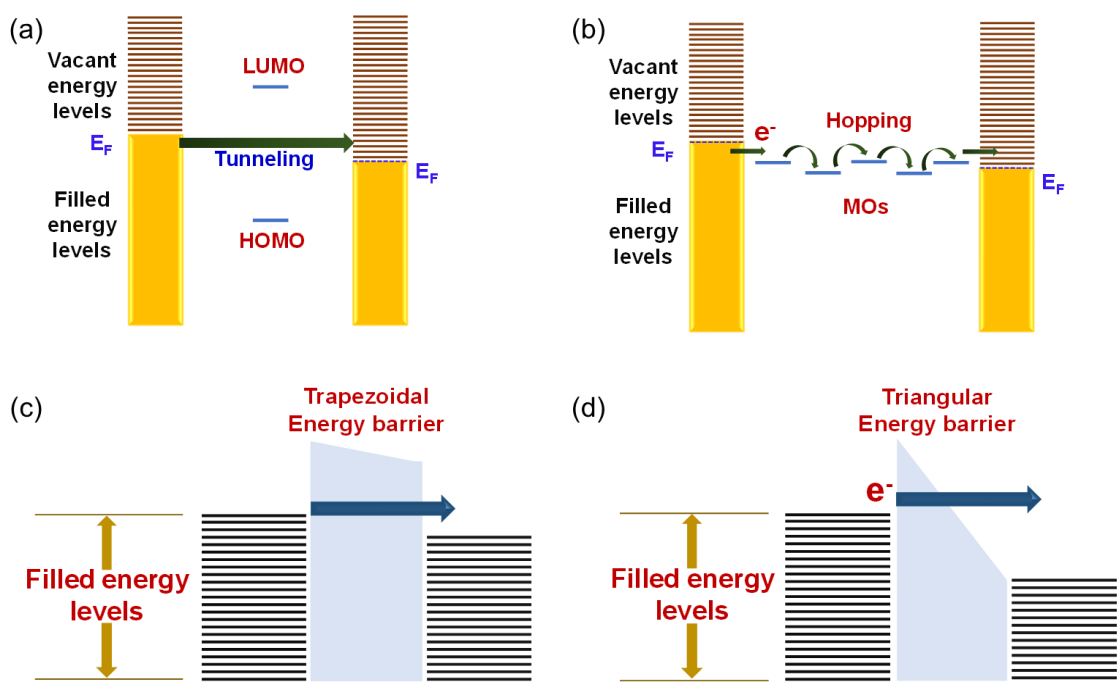

**Figure 3:** Illustration of charge transport mechanism through molecular junctions a) Tunneling, b) Hopping, C) direct Tunneling and, d) Fowler-Nordheim Tunneling.

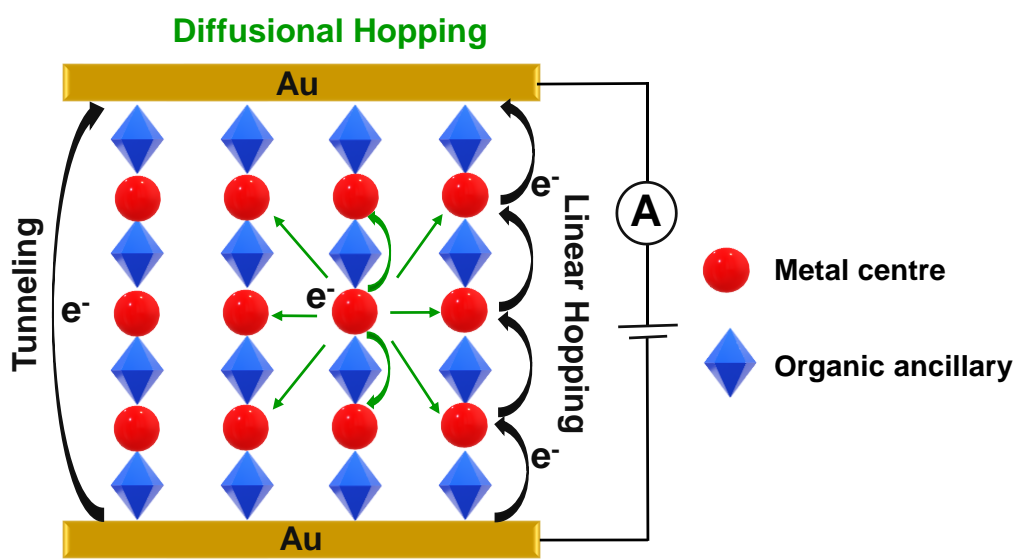

**Figure 4:** Schematic representation of two different types of hopping in molecular electronic device.

Molecular electronics can be classified as single-molecule or multi-molecule electronics and in the following sections we will delve deeper into each category using relevant examples.

### 3.1 Single-molecule (Electrode-Molecule-Electrode) charge transport

Single-molecule electronics stand at the forefront of molecular device advancement. Studies focusing on the conduction properties of these systems have revealed their inherent switching behavior—a critical function susceptible to external stimuli modulation. In these systems, switching manifests as a transition from a low to a high current state achieved by manipulating the electric field. This capability holds significant importance as it finds versatile applications in signal processing, logical data manipulation, and storage functionalities within various technological domains. The dependence of conductance on the energy levels of the electrode ($E_F$) and molecule ($E_{MO}$) has already been established. Ting and coworkers demonstrated the switching property by controlling $E_F$, employing extended metal atom chains (EMACs) for their study (**Figure 5a**).[27] Their approach involved applying potential to the working electrode ($E_{wk}$) concerning the reference electrode to modulate $E_{MO}$.

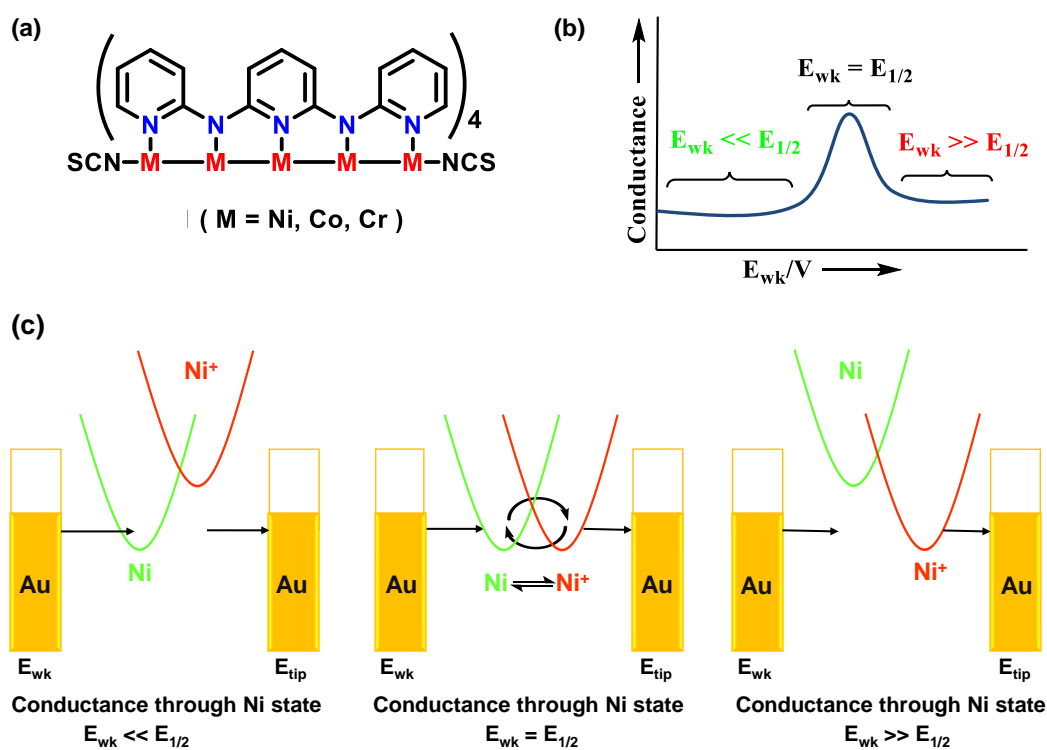

**Figure 5.** (a) Schematic representation of the extended metal atom chains supported by four oligo-α-pyridylamido anions. (b) Depiction of switching behavior at $E_{1/2}$. (c) Depiction of electron transport through an electrode/Ni$_5$ EMAC/electrode junction under EC control.[27]

Notably, the oxidation of the metal occurred at $E_{wk}$, inducing changes in conductance. Specifically, an increase in conductance was observed in the case of Ni-based molecules, while the conductance remained nearly constant for Co-based electrodes. This intriguing discrepancy was explicated by examining the alteration in bond order consequent to oxidation. Upon electron removal from the Ni-based molecule, originating from the antibonding molecular orbital (MO), the bond order decreased. Conversely, in the case of the Co-based molecule, electron extraction was from the nonbonding MO, thereby preserving the bond order. Correspondingly, the conductance aligned with these changes. The determination of $E_{1/2}$ was accomplished through cyclic voltammetry measurements. Plotting conductance against $E_{wk}$ variations revealed a distinct peak at $E_{1/2}$ (**Figure 5b**). This observed switching behavior was elucidated by the presence of a co-tunneling process, where both neutral and oxidized states reached equilibrium at $E_{1/2}$, consequently facilitating co-tunneling (**Figure 5c**).

In another study conducted by Harzmann *et al.* the concept of spin-crossover to achieve the desired switching behavior was introduced.[28] Their experimentation involved a molecule based on two terpyridine ligands, where one ligand contained two thiol groups enabling junction formation with gold via the MCBJ method. The other ligand featured push-pull substituents on each side. In the ground state, the crystal field interaction was robust, forming a low spin complex where both terpyridine ligands remained perpendicular to each other. However, under the influence of an electric field surpassing a certain threshold value, the second terpyridine ligand underwent bending (**Figure 6**). Consequently, the crystal field interaction decreased, leading to a high spin state. It was hypothesized that this high spin state exhibited higher conductivity than the low spin state. Experimental verification was achieved through *I-V* curves, which demonstrated hysteresis behavior. Interestingly, this behavior was uniquely observed in Fe-based ligands, whereas Ru-based ligands did not exhibit similar conductive switching behavior.

Schwarz *et al.* also explored the switching behavior of similar systems, employing Fe, Ru, and Mo-based ligands (**Figure 7a**) and conducted MCBJ experiments at 50 K.[29] Their analysis of the I-V curve displayed substantial hysteresis for Mo-based ligands, while Fe and Ru-based ligands exhibited minimal hysteresis, confined to specific regions (**Figure 7b**). Remarkably, in the case of Mo-based ligands, the transition from high to low current within the hysteresis curve was notably high, reaching values as high as 1000. Investigations unveiled that both hopping and tunneling mechanisms operate coherently within these systems. Consequently, DFT calculations were undertaken, encompassing both mechanisms. The results elucidated that the extensive hysteresis observed in Mo-based ligands stemmed from its substantial contribution from the charge hopping state. Furthermore, the calculations indicated the presence of a spin-localized molecular orbital in Mo, identified as a crucial factor contributing to the observed significant switching behavior.

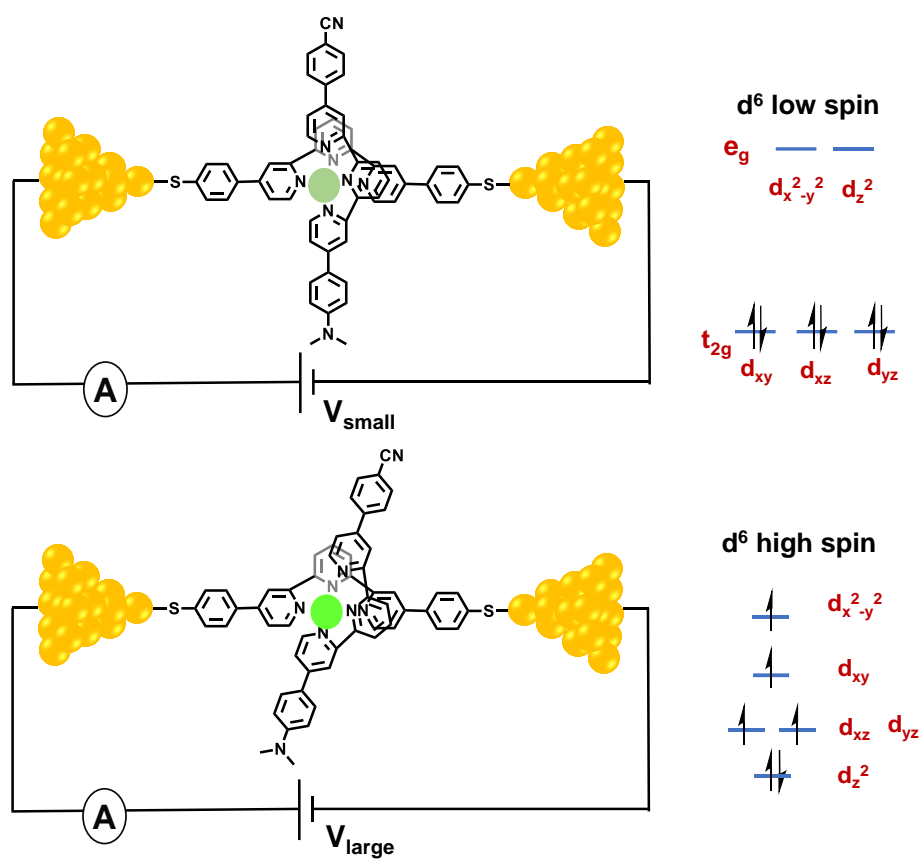

**Figure 6.** Depiction of switching behavior by changing from low spin to high spin Fe$^{II}$ complex. Reproduced with permission from ref (10). Reproduced with permission from ref (28). Copyright 2015 Wiley Publications.

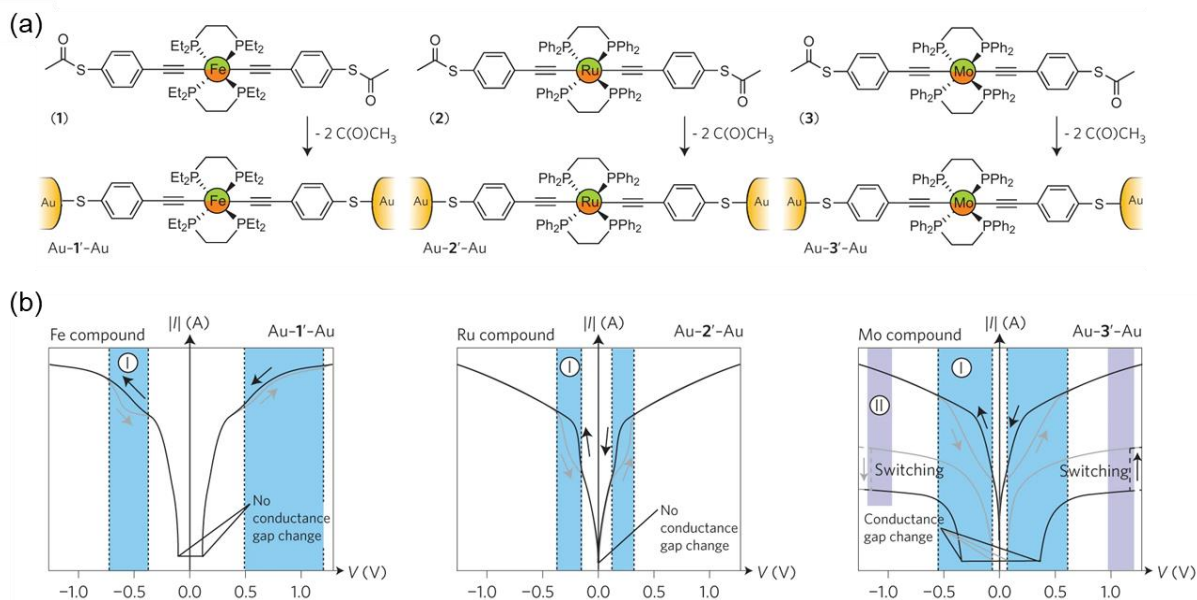

**Figure 7.** (a) Mononuclear compounds of with Fe, Ru and Mo metal centres. Au–**1′**–Au, Au–**2′**–Au and Au–**3′**–Au are representative transport junctions. (b) Depiction of the two types of hysteresis with continuous type I and abrupt switching type II. Reproduced with permission from ref (11). Copyright Springer Nature.

Ozawa and coworkers fabricated Au|single Ru complex|Au junctions aiming to investigate the influence of ancillary groups on electrical conductance (**Figure 8a**).[30] Employing the canning tunneling microscopy-break junction technique, they generated attenuation plots. The outcomes unveiled that the conductance observed with terpyridine-based ancillary groups surpassed that of 2,3-dihydrobenzo[b]thiophene-based ancillary groups. This observation firmly established that the nature of anchor groups played a pivotal role in controlling the strength of electronic coupling to the electrodes and in positioning the energy levels implicated in the transport mechanism. Additionally, the study discerned that the predominant transport mechanism was phase-coherent tunneling, resulting in a decrease in conductance as the length increased (**Figure 8b-e**). To fortify their findings, DFT calculations were executed. Calculations of the Fermi energy for all junctions were conducted and correlated with the experimentally obtained data. The computational modeling of the junctions not only corroborated the observed experimental results but also predicted that the charge transport predominantly involved LUMO-dominated transport mechanisms.

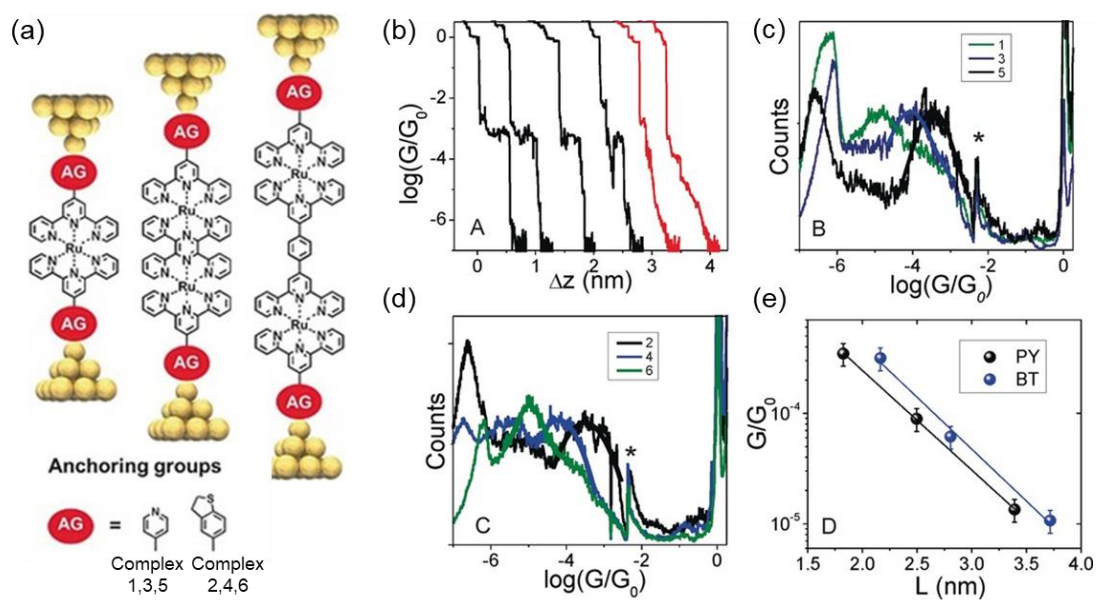

**Figure 8.** (a) Single molecule junctions with Ru-complex sandwiched between Au electrodes containing pyridine or dihydrobenzo[*b*]thiophene as the anchoring group. (b) Conductance-distance traces of complex **1** with (red) and without junction (black). 1D conductance histograms of (c) complex **1**, **3**, **5** and (d) complex **2**, **4**, **6**) terminated ruthenium complex molecular wires. (e) Probable single junction conductance values for two kinds of Ru-complexes. Reproduced with permission from ref (30). Copyright 2016 Springer Nature.

Atesci and coworkers presented a compelling advancement in the field, demonstrating the ability to control switching behavior through humidity variations.[31] Their experiment involved three distinct molecular structures (ITO|molecule|ITO). Interestingly, in the cases of molecules depicted in **figures 9a** and **9b**, rectification remained consistent and close to unity. However, in the structure illustrated in **figure 9c**, a notable shift in rectification was observed under differing humidity conditions. Specifically, under humid conditions (~60% humidity), rectification soared to 1000 for V > 0.7, whereas in dry conditions (~5% humidity), rectification remained close to unity (**Figure 9d,e**). The elucidation of these observations was achieved through DFT. In dry conditions, the HOMO and HOMO-1 combine to form two degenerate localized Molecular Orbitals, with one centered around the Ru atom and the other around the Ru substrate. When a bias is applied, these orbitals symmetrically shift, reducing the transmission. However, in humid conditions, water molecules adhere to the tip due to high hydrophilicity and capillary effects. These water molecules hydrate the counterions, displacing them by 0.5 Å and altering the local electrostatic environment. Consequently, this displacement shifts the localized molecular orbitals (LMOs) and breaks the symmetry. Computational calculations demonstrated that a positive bias aligns these changes in LMOs, thereby enhancing transmission, while a negative bias further separates them, diminishing transmission. These alterations in LMOs resulted in enhanced rectification under humid conditions, illustrating the significant impact of humidity on the switching behavior in these molecular structures.

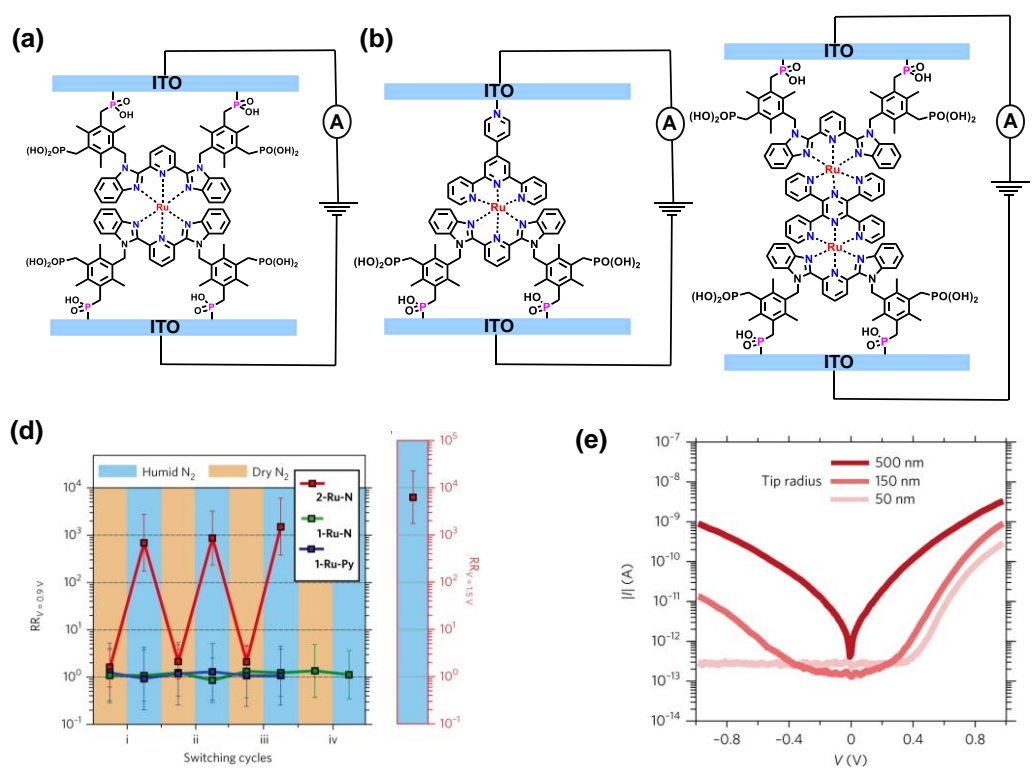

**Figure 9.** (a-c) Ruthenium complexes studied for humidity-dependent switching rectification ratio. (d) Rectification ration ($V = 0.9$ V) for Ru-complexes SAMs in dry and humid conditions. (e) Effect of different tip radii on *I-V* curves for 2-Ru-N SAMs at high humidity. Reproduced with permission from ref (31). Copyright 2018 Springer Nature.

Goswami *et al.* achieved a significant breakthrough in nanometer-scale uniform conductance switching within molecular memristors by utilizing two azo-aromatic complexes: $[Ru(L_1)_3](PF_6)_2$ ($L_1$ = 2-(phenylazo)pyridine) and system-B: $[Ru(L_2)_2](PF_6)_2$ ($L_2$ = 2,6-bis(phenylazo)pyridine).[32] Their experimental setup involved layering these materials through spin-coating, producing a thickness ranging from approximately 15–70 nm, onto an epitaxial ITO film approximately 60 nm thick, fabricated via pulsed laser deposition. In this configuration, the ITO thin film functioned as the bottom electrode, while the c-AFM tip served as the top electrode. The molecular redox mechanism enabled a remarkable uniformity of 100% switching across the entire device area, maintaining consistency even under challenging conditions, such as after $10^4$ sweep cycles and exposure to 80 °C for 2h. To delve deeper into the underlying mechanisms, they employed a nano-Raman probe-based tip-enhanced Raman spectroscopy technique. This innovative approach facilitated the deterministic identification of the molecular fingerprint and its spatial distribution within each conductance state, achieving an impressive sub-10 nm spatial resolution. This breakthrough provided detailed insights into the molecular-level behavior governing the conductance switching process, offering promising prospects for advancements in nanoelectronics and the field of molecular device engineering.

Whitesides and coworkers fabricated Au/BIPY-MCl$_2$/EGaIn junctions to gain insights into the charge transport mechanism of metal-bipyridyl complexes using SAMs of alkanethiols terminated with bipyridine moiety complexed to metal ions of 3d series (M= Cr, Mn, Fe, Co, Ni, Cu) (**Figure 10**).[33] They observed that the Mn, Fe, Co, and Ni complex shows rectification while I-V curves are almost symmetric with bias polarity for Cr and Cu complexes. Rectification is defined as the preferred flow of charge carriers in one direction when bias is applied to a molecular junction or as an asymmetry in the current-voltage characteristics over a certain potential range.[34,35] To figure out the exceptional behavior of Cr-BIPY and Cu-BIPY-based MJs, UPS measurements, and DFT calculations were done to estimate the frontier molecular orbital (FMO) energy levels. In molecular junctions, the alignment of frontier molecular orbitals with the fermi level of the electrodes plays a crucial role in charge transport.[36–38] As the MO which is close to the fermi level of the electrode contributes more to charge conduction because of the low barrier height. Therefore, Charge transport in Au/BIPY-MCl$_2$/EGaIn junctions is presumed to be favorable via LUMO during positive bias and Via HOMO during negative bias polarity. At +1 V the LUMO level of all the above-mentioned cases is in the transmission window but when a negative bias is applied i.e., -1 V only the HOMO levels of Cr-BIPY and Cu-BIPY terminated thiols lie in the transmission regime. It is the inaccessibility of the HOMO in the case of Mn, Fe, Co, and Ni complex that reduces the transmission probability, therefore, the current density, and hence accounts for the observed rectification. Dependence of current density over the temperature in the case of Cu-BIPY MJs during either bias polarity suggests the involvement of FMO levels in charge conduction while in the case of Co-BIPY MJ when a negative bias is applied, current density remains almost constant over a wide temperature range depicting the inaccessibility of HOMO and matches well with the theoretical prediction. This alignment of the FMO energy levels to the fermi level of the electrodes brings a change in the charge transport mechanism. FN plots reveal that the Co-BIPY MJs follow Fowler-Nordheim tunneling while charge transport in Cu-BIPY MJs is via direct tunneling.

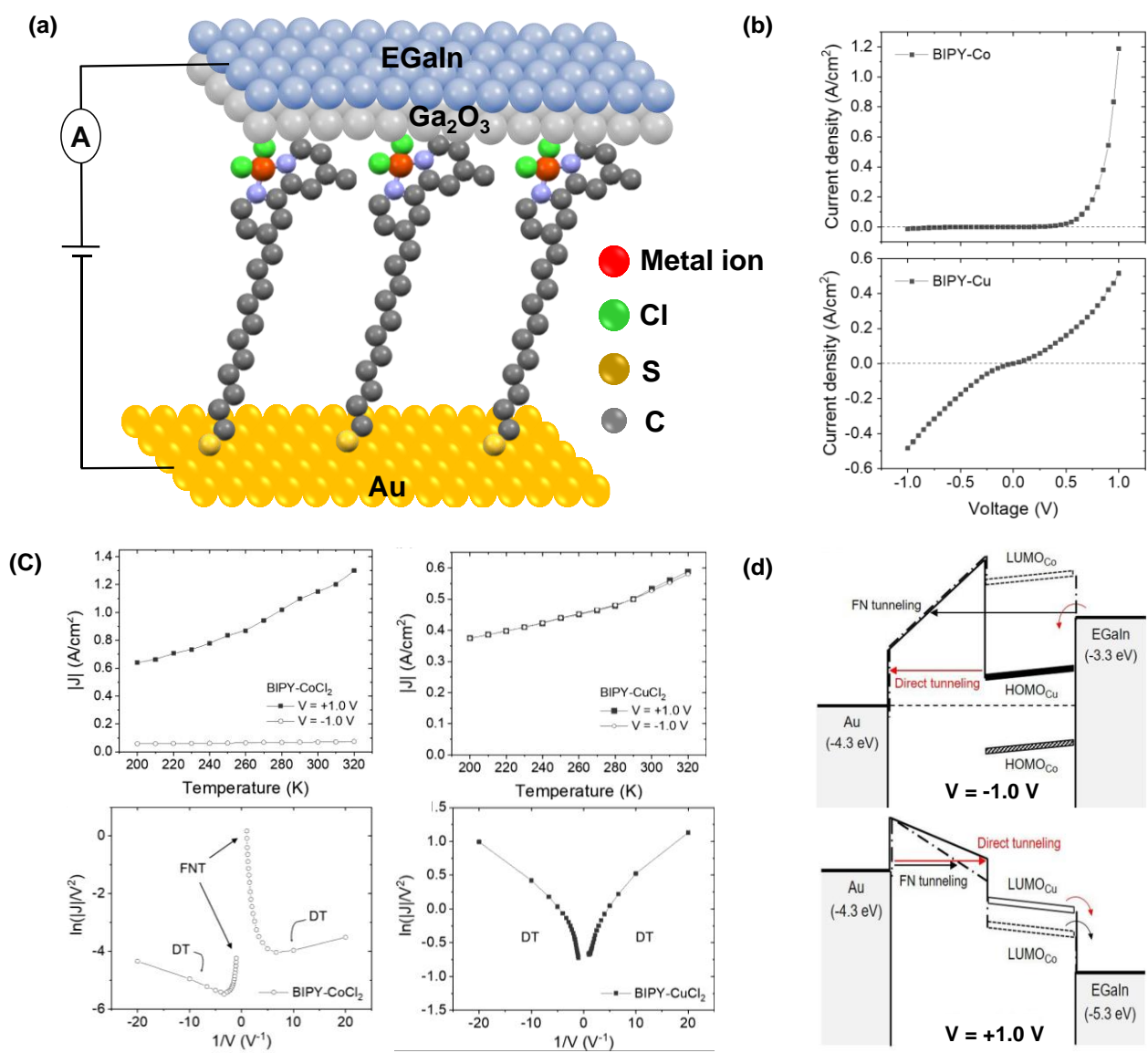

**Figure 10.** Schematics for Au/BIPY-MCl$_2$/EGaIn MJ, (b) J-V characteristics for BIPY-Co and BIPY-Cu MJs, (C) temperature dependent measurements and FN Plots for Au/BIPY-Co/EGaIn and Au/BIPY-Cu/EGaIn MJs, (d) Plausible mechanism of charge transport in these MJs. Reproduced with permission from ref (33). Copyright 2021 American Chemical Society.

### 3.2 Multicentre Molecular Charge Transport

The broad scope of applications within single molecular electronics faces constraints due to the confined space between two electrodes. To overcome this limitation, a pivotal focus lies in investigating extended wires, characterized by numerous metal centers and controllable lengths. This class of structures offers superior manipulation and adjustability concerning their inherent properties. Moreover, a critical necessity emerges for increased spacing between junctions to enable seamless integration into various devices. Consequently, a comprehensive exploration of the charge transport phenomena within these intricate systems becomes imperative.

Rampi and colleagues significantly contributed insights into transport mechanisms within multilayered transition metal junctions.[39] Their work involved the synthesis of Fe and Co-based terpyridine SAMs on an Au surface, extending up to 40 nm in length. Utilizing mercury-drop electrode, they established top contacts, forming Au-[Fe(tpy)$_2$]$_n$-Hg and Au-[Co(tpy)$_2$]$_n$-Hg junctions (**Figure 11a**). Remarkably, these junctions exhibited exceptional stability and demonstrated conducting properties over considerable distances. Analysis through attenuation plots revealed that both wires operated via a hopping mechanism for charge transport. Further investigation delineated the nature of hopping as intramolecular directional hopping (**Figure 11b-d**). Similarly, Karipidou *et al.* employed analogous complexes to synthesize Au-[Fe(tpy)$_2$]$_n$-Au junctions, spanning up to 30 nm in length.[40] In this case, the top junction was formed through thermal deposition of gold. Their detailed modeling efforts elucidated the specific nature of hopping mechanisms within this system. Notably, due to its capacity to withstand high temperature and superior mechanical stability, their experimental observations aligned with the Richardson–Schottky Ansatz model for the thermionic emission model asshown in equation 3.

$$J = A^* T^2 e^{\left(\frac{-\varphi_b}{kT}\right)} e^{\left(\frac{\sqrt{\frac{q^3 V}{4\pi\varepsilon_o}}}{kT}\right)} \ldots\ldots(3)$$

where $A^*$ is the effective Richardson constant, $\varphi_b$ is the energy barrier for hole/electron injection (Schottky barrier), $\varepsilon$ the dielectric constant, $\varepsilon_o$ the permittivity of vacuum, $k$ the Boltzmann constant, and *q* the electronic charge. The results obtained by experiments were also verified using computational methods. The study revealed that charge transport occurs through electron hopping both along and off the wire. Additionally, it identified a Schottky barrier with a $\varphi_b$ value of 0.72 eV, limiting the charge injection process.

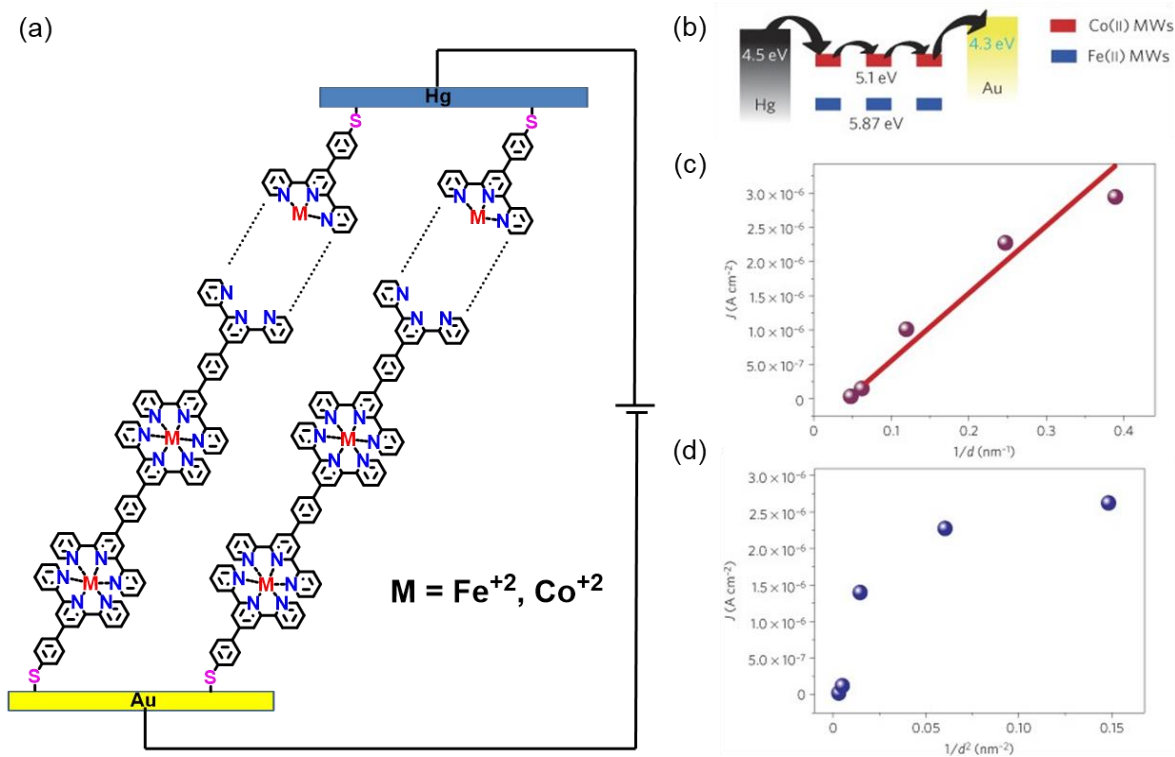

**Figure 11**. (a) Representation of Au-[Fe(tpy)$_2$]$_n$-Hg and Au-[Co(tpy)$_2$]$_n$-Hg based molecular wire. (b) Corresponding energy diagram for junctions. (c) Plots of current density (*J*) versus respectively $1/d$ and $1/d^2$ for the Fe$^{2+}$-based molecular wire. Reproduced with permission from ref (39). Copyright Springer Nature.

Nguyen et al. recently presented compelling findings on charge transport within multilayered molecules, primarily comparing organometallic molecular junctions (MJs) with organic molecular Junctions.[41] Notably, the investigation focused on transition metal-based MJs, employing cobalt-based complex [Co(tpy)$_2$](PF$_6$)$_2$ and ruthenium-based complex [Ru(bpy)$_3$](PF$_6$)$_2$. These MJs were established through in situ diazonium ion generation on Au and direct chemical deposition on Ti/Au layers, resulting in Au|molecules|Ti|Au configurations (**Figure 12a-b**). Analysis *via* an attenuation plot, correlating the log of current density with molecular thickness variations between the electrodes, unveiled intriguing observations specific to Ru-based MJs (**Figure 12c**). Notably, two distinct β values were identified, with a transition occurring at a specific length, d$_{trans}$ = 3.5 nm. Below d$_{trans}$,

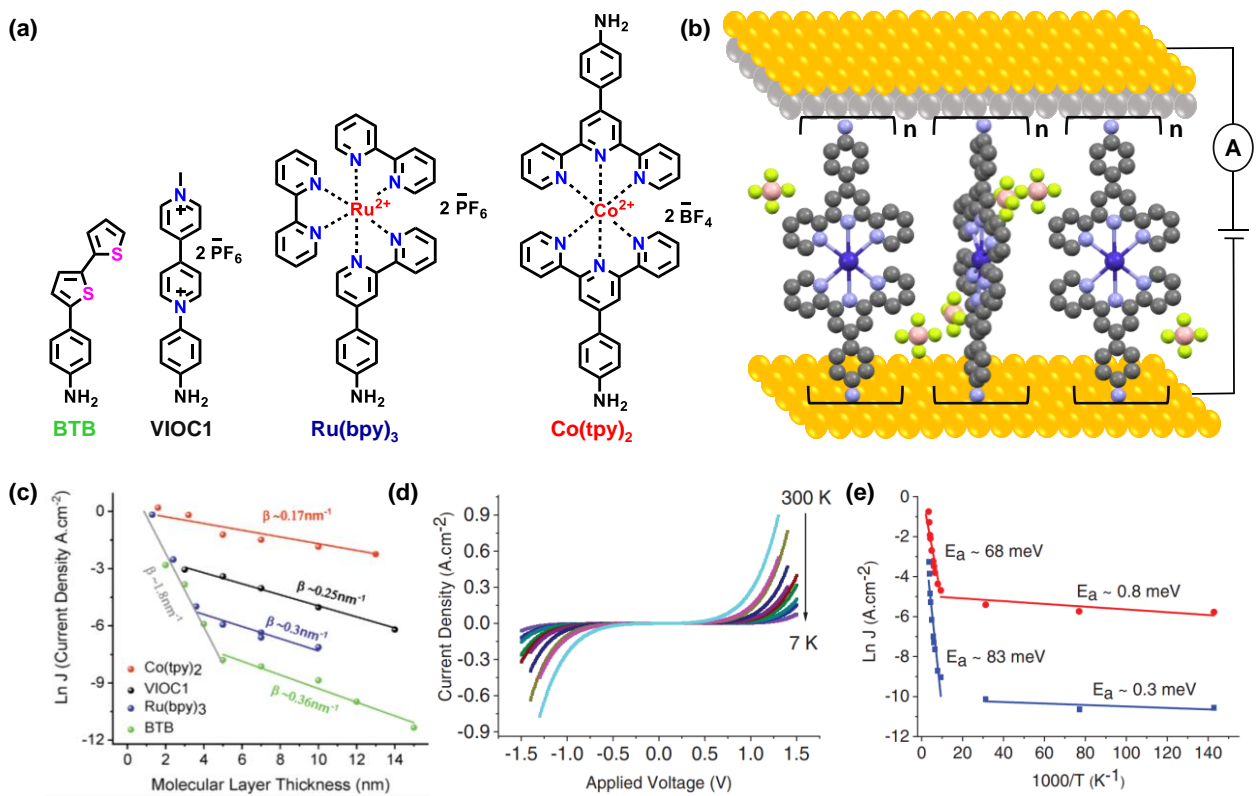

**Figure 12:** (a) Structures of molecules. (b) Co(tpy)$_2$ molecule-based molecular junction. (b) Attenuation plots taken at 1 V of MJs for different molecular units. (d) Temperature dependent J-V overlay of [Co(tpy)$_2$] 7 nm MJs. e) Arrhenius plot taken at 0.5 V (blue) and 1 V (red). Reproduced with permission from ref (41). Copyright 2020 Wiley Publications.

$β$= 1.8 nm$^{-1}$, indicating non-resonant direct tunneling, while above d$_{trans}$, β = 0.3 nm$^{-1}$, signifying transport mechanisms involving field ionization, multistep tunneling, and associated redox events. The attenuation plot analysis of Co-based molecular junctions exhibited a consistent β value, indicating characteristic long-range charge transport behavior, notably registering one of the lowest values at 0.17 nm$^{-1}$ within this system. To comprehend the transport phenomena in these junctions, two key studies were conducted. Firstly, an exploration of the energy levels of electrodes and the molecule revealed a convergence and resonance of all energy levels upon junction formation, favoring resonant tunneling despite slight initial disparities in their free states. Secondly, temperature's effect on the system was investigated. Analysis of the current density plotted against the electric field for a 7 nm molecular junction unveiled activation energy findings (**Figure 12d-e**): above 100 K, it ranged between 80 and 60 meV, while below 100 K, it diminished drastically to approximately 0.3 meV. As a result, the authors concluded that below 150 K, resonant tunneling predominates. However, at room temperature (or above 150 K), two mechanisms become viable:

resonant tunneling and interchain hopping, contingent upon redox events, both of which are mutually exclusive.

In a separate study by the same group, similar observations were made regarding Co and Ru-based terpyridine junctions, providing complementary insights.[42] They employed STM to analyze the transport mechanisms in Au-[Co(tpy$_2$)]$_n$-Au, Au-[Ru(tpy)$_2$]$_n$-Au, and Au-[Ru(bpy)$_3$]$_n$-Au junctions (**Figure 13a-b**), a notably rare assessment for this technique, spanning lengths from 2 nm to 8 nm. The grafting of complexes onto the first Au was achieved through the electroreduction of the respective diazo complexes, while the second junction was established using the STM-BJ technique. Conductance measurements were obtained using the STM-BJ technique, and $\beta$ values were derived from the attenuation plots (**Figure 13c**). For Au-[Co(tpy$_2$)]$_n$-Au, an exceptionally low $\beta$ value (~0.19 nm$^{-1}$) was observed for both high conductance and low conductance states, accompanied by high conductance (~10$^{-3}$ G$_o$). These findings support the proposal of a resonant tunneling mechanism.

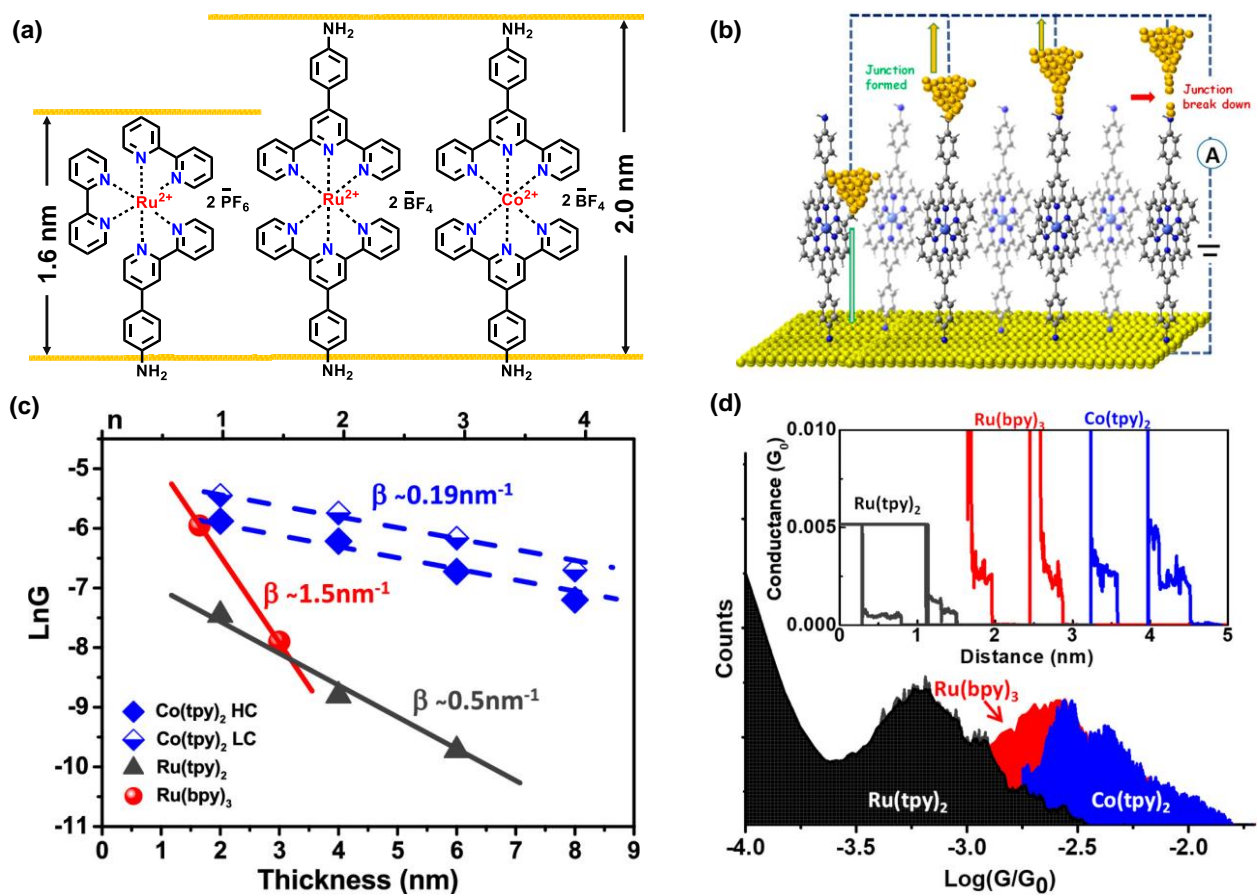

**Figure 13:** (a-b) Schematic diagram representing Au-[Fe(tpy)$_2$]$_n$-Au, Au-[Co(tpy)$_2$]$_n$-Au and Au-[Fe(bpy)$_3$]$_n$-Au. (c) Attenuation factor $\beta$ vs thickness dependence various MJs, (d) conductance histogram of MJs. Reproduced with permission from ref (42). Copyright 2020 American Chemical Society.

Regarding Au-[Ru(tpy)$_2$]$_n$-Au, a lower conductance value (~$10^{-4}$ G$_o$) compared to Au-[Co(tpy$_2$)]$_n$-Au was noted, along with a $\beta$ value of 0.5 nm$^{-1}$. The relatively higher $\beta$ value suggests the dominance of interchain hopping as the primary mechanism. In the case of Au-[Ru(bpy)$_3$]$_n$-Au, the higher $\beta$ value (1.5 nm$^{-1}$) alongside substantial conductance that decreases rapidly with length implies a non-resonant tunneling mechanism. The group predicted a transition from tunneling to hopping mechanisms around a length of approximately 3 nm for Ru-based junctions. These observations provide valuable insights into the transitioning mechanisms governing charge transport behavior in these complex molecular junctions.

Sachan and Mondal conducted the preparation of both homostructured Fe$^{2+}$/Co$^{2+}$-bis-terpyridine (bis-tpy) oligomers, as well as heterostructures, on ITO-coated glass substrates utilizing the LbL technique (**Figure 14a**).[43] Their methodology involved initially grafting the terpyridine template layer electrochemically using a diazo precursor. Subsequently, a coordination-driven strategy was employed to fashion an assembly of ITO|tpy-[Fe(bis-tpy)]$_n$ or ITO|tpy-[Co(bis-tpy)]$_n$ (where n = 2-8). They constructed a Molecular Junction (MJ) comprised of ITO/tpy-[Fe(bis-tpy)]$_5$-gel electrolyte-[Co(bis-tpy)]$_5$-tpy/ITO, which intriguingly exhibited no conduction within the ±2 V window in the *J-V* curve (**Figure 14b**). However, asymmetrical conduction was observed in both directions beyond this voltage window. This asymmetrical conduction phenomenon was attributed to the distinct energy levels of $E_F$–LUMO/$E_F$–HOMO for $Fe^{2+}$ and $Co^{2+}$ molecular assemblies. The transport of electrons was suggested to occur from the Fermi level of the ITO to the LUMO of the $Fe^{2+}$/$Co^{2+}$-bis-terpyridine compounds, given that this energy barrier is significantly lower than that of EF–HOMO, measuring approximately 2.43 eV and 2.54 eV for $Fe^{2+}$ and $Co^{2+}$-terpyridine thin films, respectively (**Figure 14c**). Moreover, Bode plot analysis derived from Electrochemical Impedance Spectroscopy (EIS) measurements showcased frequency-dependent capacitance and resistance behavior of the Molecular Junction (**Figure 14d**). The capacitance exhibited maximal values in the lower frequency range (10 to $10^3$ Hz) and at a phase angle (θ) of -75º, while minimizing at higher frequencies with a phase angle of θ = 0º.

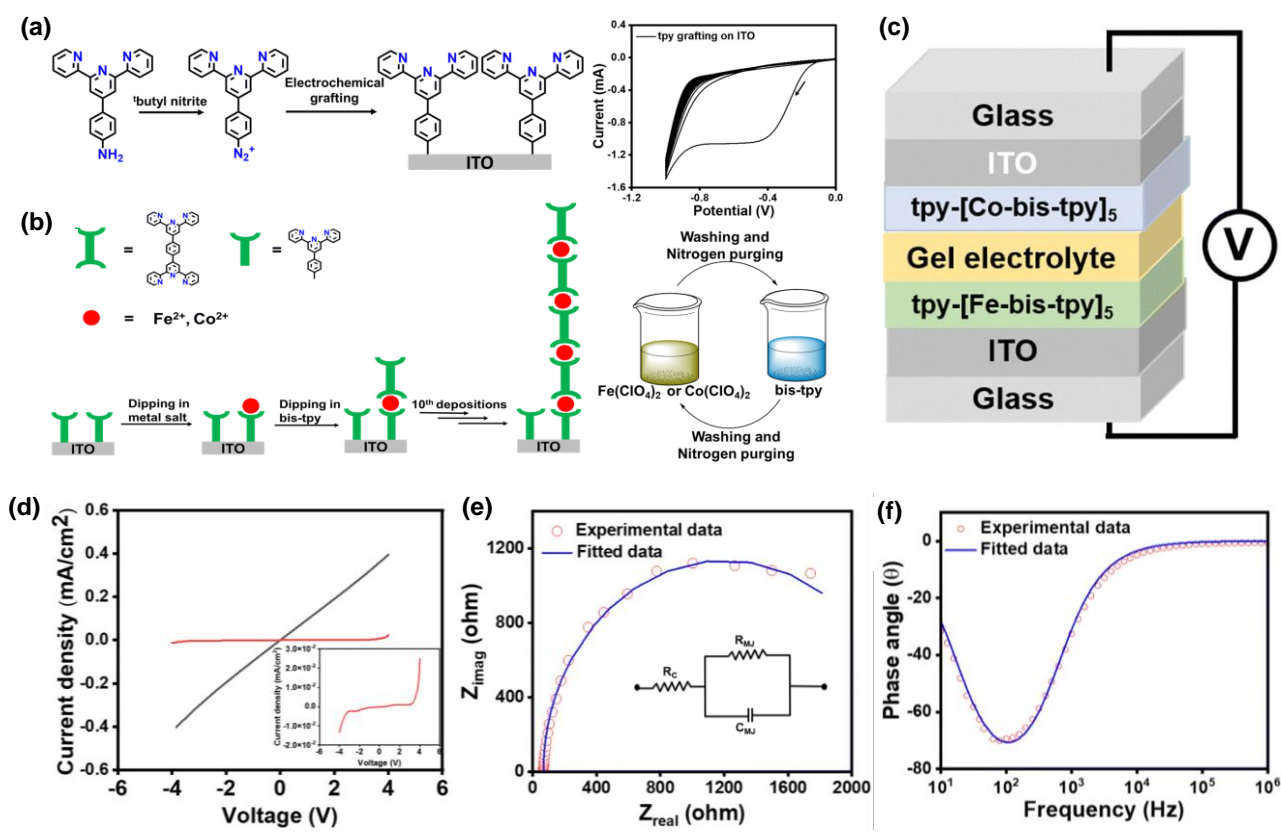

**Figure 14**. (a-b) Schematic illustration of the layer-by-layer oligomer formation employing electrochemical grafting technique (c) Schematic description of MJ (d) J-V plot of the MJ (red) and a reference junction (black). (e) Nyquist plot for the MJ fitted with Randles-circuit (f) Bodes plot for the MJ showing frequency and phase angle dependencies. Reproduced with permission from ref (43). Copyright 2022 Royal Society of Chemistry.

Vitale *et al.* synthesized a novel molecular junction having two $Fe^{2+}$ and $Ru^{2+}$ centers.[44] The grafting of molecules was done on zirconyl phosphate-covered fluorine-doped tin oxide (ZP-TFO) through layer-by-layer method and the top contact was made from the eutectic Ga-In alloy. For n = 1, tunnelling was observed to be the transport mechanism from attenuation plot. But the length was well on the transition regime. So, from n > 1, the transport mechanism observed was space-charge limited conduction. This was confirmed from the linear plot of *J vs* $d^{-3}$ where *d* is the distance between the electrodes.

Ruthenium-polypyridyl complexes are well-known for their one-electron reversible redox nature and their photophysical, catalytic, sensing, and biological properties are extensively studied.[45–50] Recently, Gupta *et al.* fabricated Ru-tris(5-amino-1,10-phenanthroline) molecular assemblies based on nanometric junctions to understand the mechanism of charge transport through ITO/Ru-N6/Al junctions.[45] Ultrathin three-dimensional molecular assemblies (4-16 nm) of Ru tris(5-amino-1,10-phenanthroline) were electrochemically grafted on patterned ITO by taking advantage of the diazotization reaction and electrochemical reduction of diazonium salts (**Figure 15**). A notable difference in current density with the thickness of the molecular junction was observed. The linear variation of Ln J with thickness d accounts for the tunneling mechanism in thinner junctions and the attenuation factor was found to be decreased from 0.79 nm$^{-1}$ to 0.70 nm$^{-1}$ while increasing the bias from +0.25 V to +1 V which depicts an efficient long-range charge conduction in these molecular junctions. The device yield of >91% shows that the electrochemical grafting technique is a promising way to make thin films of molecular assemblies. To elucidate the mechanism of charge transport effect of temperature and electric field on J-V curves was examined. Temperature dependent J-V measurements, FN and the Poole-Frankel plots depict the field emission mechanism for charge transport in thicker junctions and resonant tunneling in thinner junctions.

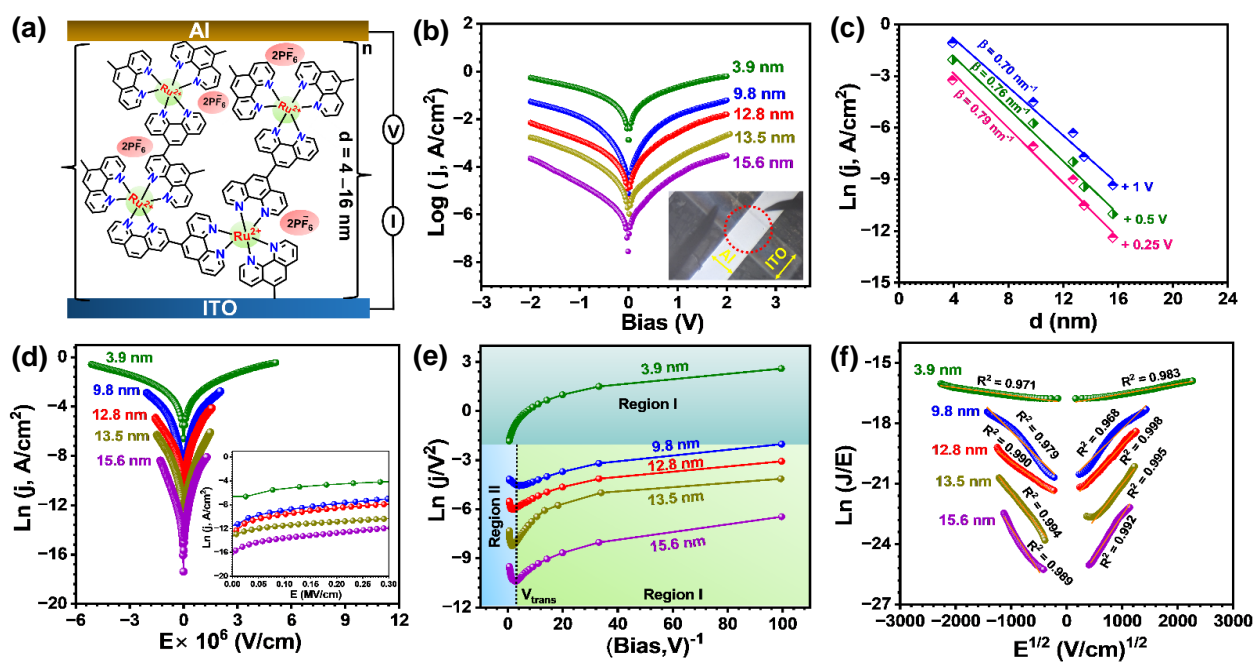

**Figure 15.** (a) Side view of two-terminal MJs with vertical stacking of ITO/[Ru(Phen)$_3$]$_{d=4\text{-}16nm}$/Al, (b) Semilog j-V curves for Ru(Phen)$_3$ MJs with the molecular thickness of 4-16 nm (inset showing an image of MJ). The J-V curves are the average of 6-7 individual MJs, (c) Ln J vs. d plot at different bias for estimating β values, (d) comparison of Ln J vs. *E* plot for different thicknesses of Ru(Phen)$_3$ MJs, and (e) FN plot and (f) PF plot for different thicknesses of Ru(Phen)$_3$ MJs. Reproduced with permission from ref (45). Copyright 2023 American Chemical Society.

## 3. Molecular Electronics: Photocurrents

Other than charge transport, metal-polypyridyl complexes have found applications in the areas of photochemistry and photophysics as well. This is primarily due to the absorbance resulting from metal to ligand charge transfer (MLCT) or ligand to metal charge transfer (LMCT), alongside their luminescent properties and reversible redox states. These attributes serve as an avenue for exploring into their luminescence behaviors. Moreover, their capacity to emit high-brightness, high-efficiency radiations at low-driving voltages positions them as interesting luminophores for both chemi- and electroluminescent devices. Their role as photosensitizing materials in dye-sensitized solar cells (DSSCs) is also well-documented. Exhibiting light absorption across a broad visible spectrum, these complexes often achieve incident photon-to-current conversion efficiencies (IPCE) exceeding 80%. Further enhancing their utility, ruthenium complexes exhibit commendable stability in both their ground and excited states, exhibiting minimal decomposition during redox processes on the photoelectrode. While their significance in DSSCs is well-established, the potential application extends beyond solar cells to optoelectronic logic devices, exemplified by their suitability in molecular switches. Notably, the tunability of the HOMO/LUMO gaps in these complexes through ligand modifications holds paramount importance, impacting the complex ability to donate, accept, or mediate electron transfer from the HOMO to the LUMO—an influential determinant in the efficacy of organic electronic devices.

Oszajca et al. explored the photoelectrochemical properties of ruthenium complexes ([Ru(tpy)Cl$_3$] (**I**), [Ru(bpy)$_2$Cl$_2$] (**II**), [Ru(bpy)$_2$(C$_2$N$_2$S$_2$)] (**III**) and [Ru(dcbpy)$_2$(C$_2$N$_2$S$_2$)] (**IV**) with the TiO$_2$ surface, creating Ti-O-Ru and Ti-N≡C- bonds (**Figure 16a**).[51] They observed differences in the absorption of **III** and **IV**, indicating strong interaction of the cyano group with TiO$_2$. The studied complexes exhibited MLCT absorption bands shifted by ~50 nm and displayed photocurrent switching effects (**Figure 16b-c**). Cases, like **I**@TiO$_2$ and **III**@TiO$_2$, mainly cathodic photocurrents occurred, whereas oxidized ruthenium species showed limited photosensitization for TiO$_2$. The materials **II**@TiO$_2$, **III**@TiO$_2$, and **IV**@TiO$_2$ displayed diverse photocurrent characteristics and photosensitization ranges. The photocurrent switching process in these materials was not associated with redox transformations but was suggested to involve competing photocathodic and photoanodic processes due to the Schottky barrier at the ITO|molecule|TiO$_2$ MJ. The molecular interlayer between TiO$_2$ and ITO was proposed to modify the Schottky barriers height through dipole interactions, influencing the photocurrent behavior. Additionally, differences in potential differences (E$_{PEPS}$ - E$_{1/2}$) (PEPS = photocurrent switching potential) implied varying energy barriers among the complexes, impacting their photocurrent responses (**Figure 16d**).

Solution phase electrochemistry of Ru(bpy)$_3$ and its derivatives have been widely explored owing to their photochemical, photophysical, and electroluminescent properties.[52,53] The first paper describing a solid-state light-emitting device based on a ruthenium phenanthroline complex was published in 1996.[54,55] Bard and coworkers fabricated a 100 nm thick Ru(bpy)$_3$-based LEDs employing spin coating technique to understand the mechanism of charge transport in solid-state LEDs.[56–58] The electroluminescence in these devices is attributed to the formation of Ru$^{3+}$ and Ru$^{1+}$ centers at positive and negative electrodes respectively, followed by migration and recombination to form $Ru^{2+*}$, which causes light emission and mobility of counter ions plays a vital role. In the same line, to understand the charge transport pathways at the nanoscale, MJs of thickness d = 5-28 nm were fabricated by Mcreery and others utilizing electrochemical reduction of diazonium salt (**Figure 17**).[59] This is achieved by probing current density and light emission as a function of bias, thickness, temperature, and electric field. In contrast to conventional organic light-emitting diodes d >50 nm, two mechanisms were found to be operative in nanometric junctions.[57] The observed positive photocurrent, linear variation of ln J vs $E^{1/2}$, and LE/Current vs $E^{1/2}$ indicate field-driven unipolar injection to be a prominent charge conduction mechanism in thinner MJs and bias-dependent bipolar injection in thicker MJs once bias exceeds 2.7 V (bandgap). Lower light emission rise time compared to previous literature reports along with the existence of two different field and bias-dependent mechanisms point towards the screening of electrodes by mobile charge carriers, which is also reflected by the measurements performed in the presence of ACN vapors.[57,58]

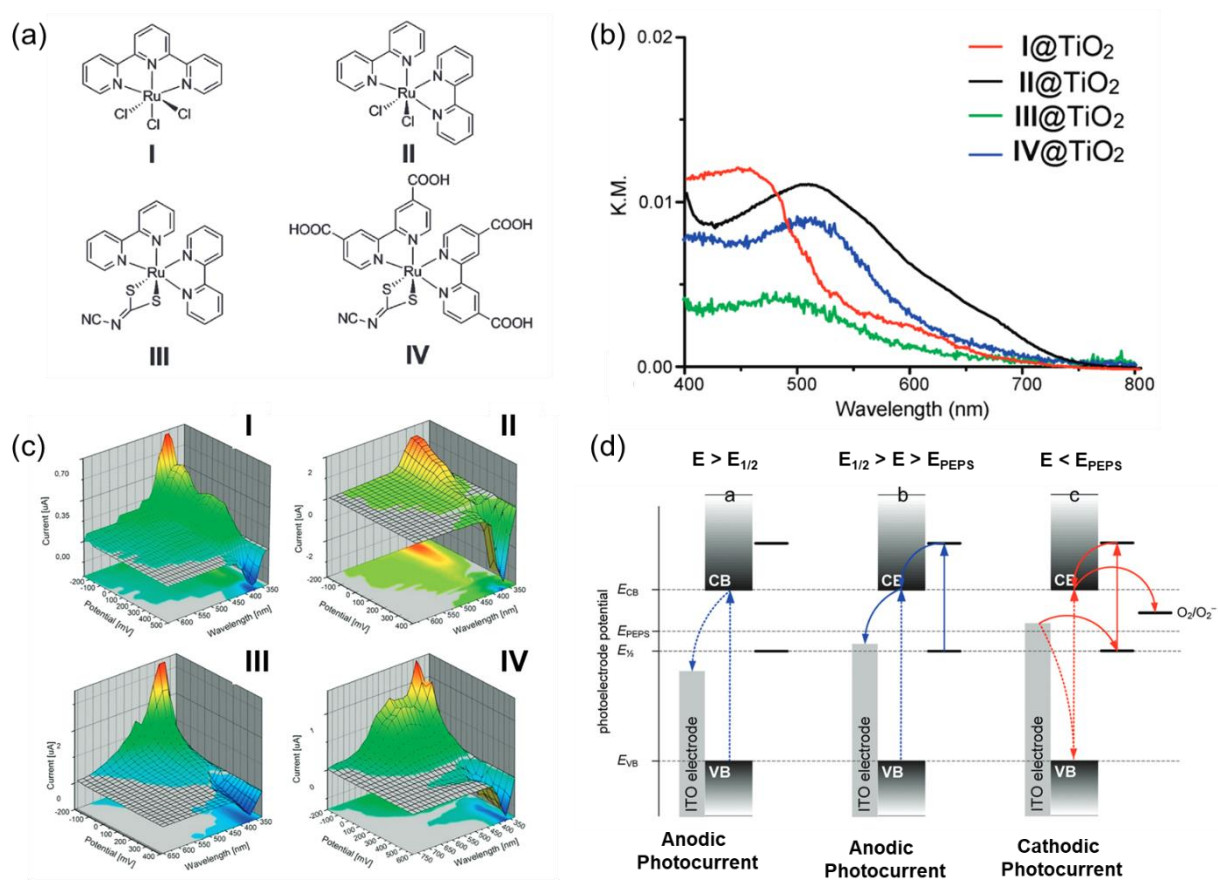

**Figure 16**: (a) Structure of investigated $Ru^{2+}$ complexes. (b) Transformed diffuse reflectance spectra of Complexes@$TiO_2$. (c) Photocurrent amplitude for photoelectrodes made of complexes@$TiO_2$. Cathodic and anodic photocurrents are depicted by red and blue regions while net-zero photocurrent is represented by grey region. (e) Mechanism of photocurrent switching in Ru-modified $TiO_2$ for anodic and cathodic photocurrent generation. Reproduced with permission from ref (51). Copyright 2011 American Chemical Society.

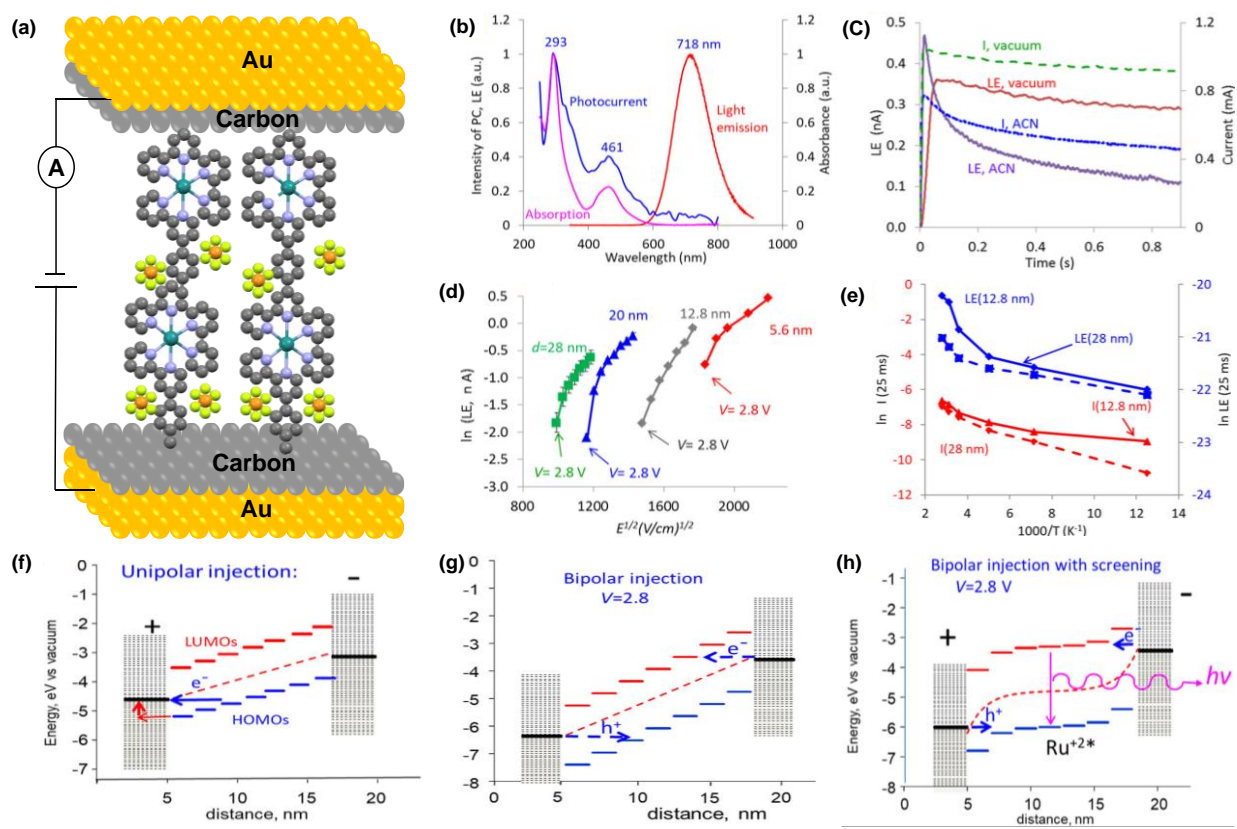

**Figure 17:** (a) schematic representation of Cr/Au/eC/Ru(bpy)$_3$/eC/Au/Cr molecular junction, (b) UV-Vis absorption, photocurrent and light emission spectra for 12.8 nm Ru(bpy)$_3$ MJ, (c) Current versus time (dashed lines) and total light emission versus time (solid lines) for bias pulses of 3.2 V at room temperature under vacuum and acetonitrile vapor, (e) Arrhenius plots, (f) energy level schematics of injection mechanisms, Unipolar injection with electron transfer from HOMO to the left electrode, with the horizontal blue arrow indicating tunneling and red arrows representing activated electron transfer. (g) Bipolar injection involving both HOMO and LUMO orbitals. (h) Bipolar mechanism assisted by electrode screening. Reproduced with permission from ref (59). Copyright 2019 American Chemical Society.

A redox couple is of paramount importance for dye regeneration in dye-sensitized solar cells (DSCs) and plays an imperative role in the photovoltage and photocurrent of the device.[60] To overcome the limitations of I$^-$/I$_3^-$ redox couple in DSCs such as low V$_{oc}$ (open-circuit potential), corrosive nature, transition metal polypyridyl complexes were explored owing to their reversible redox properties and thermodynamic stability attributed to chelate effect.[61,62] Cobalt-polypyridyl-based DSCs were shown to exhibit solar-to-electric power conversion efficiency of 14% with an improved open circuit potential up to 1 V which is limited by the slow mass transport and preorganization energy involved during redox event $d^7$(high spin) to $d^6$(low-spin) in cobalt-polypyridyls.[63] Inspired by the efficient electron-transfer nature of blue-copper proteins, copper-based redox electrolytes were intensively studied[64] and $V_{oc}$ increased over 1 V alongside PCE of 15.2 % under 1 sun illumination using [Cu(tmby)$_2$]$^+$/[Cu(tmby)$_2$]$^{2+}$ redox shuttle ascribed to its low inner sphere reorganization energies and high redox potential[65,66]. It is appealing to tune the redox potential of copper-based redox couple to further enhance the performance of DSCs with already-developed dyes.[67] For this purpose, ligand electronic structure is tailored, and preorganized tetradentate, pentadentate, and hexadentate ligands have been developed for faster electron-transfer kinetics. In a recent study by Devdass *et.al.* Cu-complex bearing preorganized hexadentate ligand has been utilized as a redox-shuttle in DSCs which operates at low voltage and has proven to provide stability to the DSC device as proximal pyridine of hexadentate ligand blocks the ligand exchanges with TBP ligand, an additive used in DSCs.[68] In Cu(bpy)PY4-based DSCs electron-transfer kinetics was found to be sluggish because of the large inner sphere reorganization energy involved (app. 1 eV per molecule) (**Figure 18**). The PCE was found to be higher under fluorescence illumination reaching a maximum of up to 15.2 %. This work provides valuable information about the design of DSCs (selection of redox system and sensitizer) for high photocurrent devices.

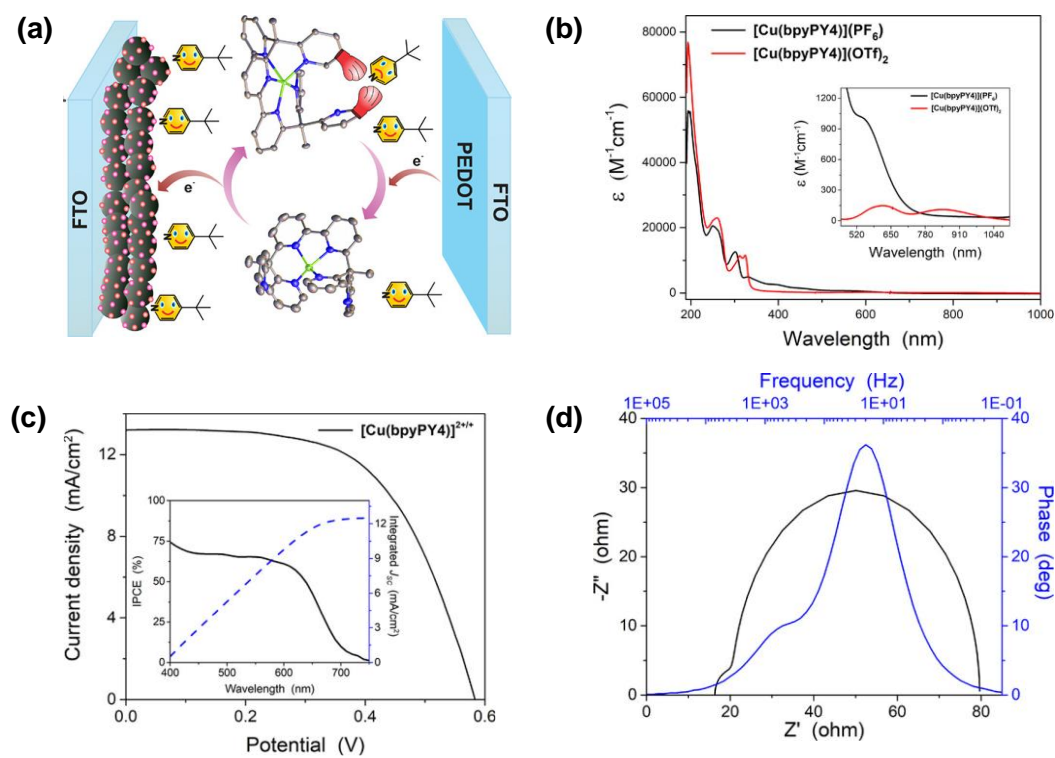

**Figure 18.** (a) schematics for the [Cu(bpyPY4)]$^{2+/+}$- based dye-sensitized solar cell, (b) UV-Vis spectra of Cu-redox shuttle in ACN, (c) J−V and IPCE (inset) curves for the [Cu(bpyPY4)]$^{2+/+}$- based DSC devices. For the IPCE, the solid black line is the IPCE response, and the dashed blue line represents the integrated photocurrent, (d) Nyquist (black) and Bode plot (blue) for DSC devices. Reproduced with permission from ref (68). Copyright 2022 American Chemical Society.

## 3. Molecular Memory

The groundbreaking work of de Silva highlighted molecules potential to process information akin to electronic systems.[69,70] Despite silicon (Si)-based memory devices market dominance due to established manufacturing and high storage density, they exhibit limitations like low endurance (approx. $10^6$ write/erase cycles), slow write speeds (1 ms/0.1 ms), and high operating voltages (>10 V). To overcome these drawbacks, diverse materials including inorganic metal oxides,[71,72] chalcogenides,[73,74] nitrides,[75] organic compounds[76,77] and polymers[78,79] have been explored for non-volatile memory devices. Classical inorganic metal oxides attract attention for their high ON/OFF ratios, endurance, and data retention. However, their complex fabrication processes and limited tunability due to rigid chemical structures hinder industry adoption. These materials replicate fundamental logic gate functions. This emulation involves intricate modulation of chemical, electrical, thermal, and optical signals within molecular frameworks, ushering in molecular-based information processing systems. Molecular memory-active materials exhibit characteristics like electrochemical activity, multiple reversible redox processes across a wide voltage window, stability in various redox states and rapid electrochemical kinetics.

Kandasamy *et al.* demonstrated the non-volatile memory capabilities of homoleptic and heteroleptic polypyridyl $Cr^{3+}$ complexes (**1-5**) (**Figure 19**).[80] Memory devices were constructed on a 300 nm $SiO_2$ wafer with an aluminum bottom electrode and subsequent deposition of two aluminum top electrodes. The *I-V* characteristics of complexes **1-4** suggested that the work function of the aluminum electrode and hole-injection through ligand-centered oxidation might be the primary pathway for these memory devices. Devices based on **1** and **2** exhibited distinct bipolar switching, that is, resistive random-access memory between high resistance states (HRS, achieved with positive voltage sweeps up to 4.6 V for **1** and 2.8 V for **2**) and low resistance states (LRS, attained with negative voltage sweeps down to -4.6 V for **1** and -2.8 V for **2**), achieving impressive ON/OFF current ratios in the range of $10^4$–$10^6$. Conversely, due to the irreversibility of oxidized complexes **3** and **4**, a Write-Once Read-Many behaviors was observed with ON/OFF ratios ranging only from 20 to 150. This study offered valuable insights into molecular design strategies for polypyridyl complexes tailored for memory device applications.

Leung et al. also reported a $Ru^{2+}$-polypyridine complex (**Figure 20a**) and a memory device with Al/$Ru^{2+}$-Complex/ITO configuration.[81] Upon initiating, the device resides in a high-resistance OFF state. During a voltage sweep from 0 to +5 V, a sharp increase in current from $10^{-8}$ to $10^{-4}$ A occurs at a switching threshold voltage of ~4 V (**Figure 20b**). This transition signifies the device switching from OFF to ON states, akin to a "writing" process in memory devices. The ON state persists during a subsequent sweep from 0 to +5 V and demonstrates stability even when a reverse bias of 0 to −5 V is applied. Notably, upon powering off, the ON state transitions back to the OFF state, resembling a static random-access memory behavior. Furthermore, another voltage bias from 0 to 5 V switches the OFF state back to the ON state at a similar threshold voltage, indicating the device's reversibility. The ON/OFF current ratio would remain over $10^4$ indicating good stability of the device (**Figure 20c**). On similar lines, Chhatwal et al. developed a $Ru^{2+}$-tpy complex (**Figure 20d**) that, when electropolymerized on ITO-coated glass (**Figure 20e**), exhibited static random access memory behavior.[82] This material offers data storage density surpassing that achieved by flip-flop and flip-flap-flop logic circuits, reaching approximately $\sim 4 \times 10^{15}$ bits/cm$^2$ and is controlled by the applied voltage and can be accessed optically (**Figure 20f**). Same group designed hetero-bimetallic polypyridyl complex based monolayer comprising two redox-active data saving bits ($Ru^{2+}$ and $Os^{2+}$) and a conductive imidazole aromatic spine.[83] They showed that by controlling the applied potential, the monolayer can show three-different redox states and can act as binary/ternary memory module.

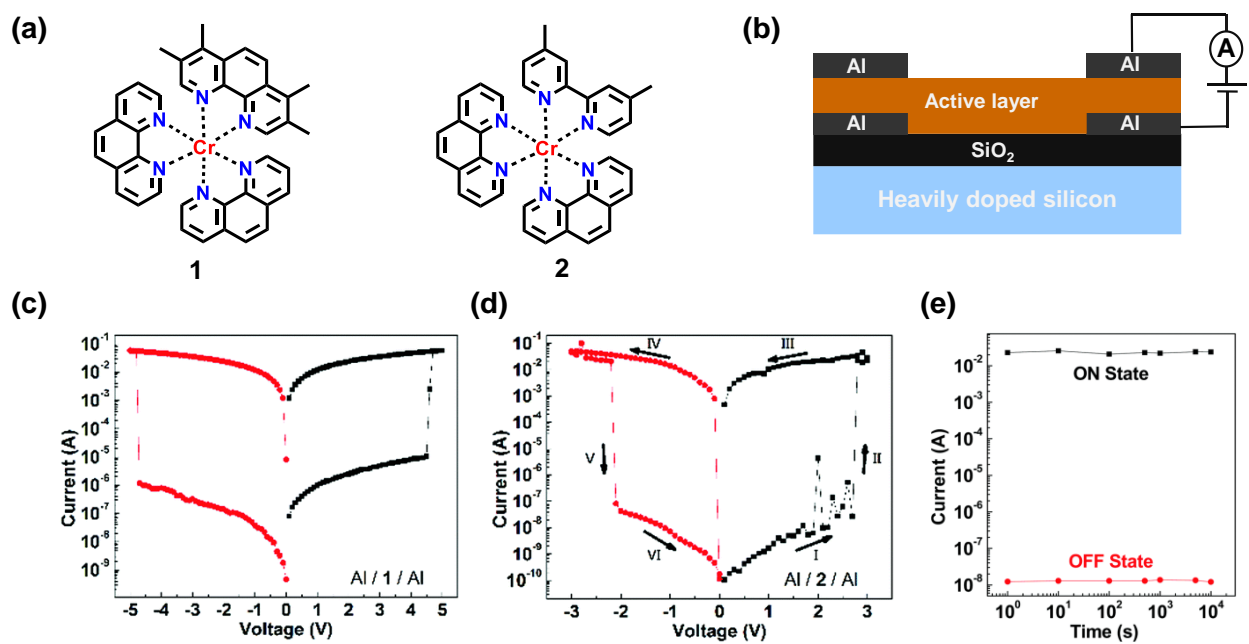

**Figure 19.** (a) Structure of $Cr^{3+}$ complexes. (b) Schematic of the memory device based on $Cr^{3+}$ complexes as the active layer. (c) I-V characteristics for Al/complex1/Al, and (d) Al/complex2/Al memory devices. (e) Bipolar switching between HSR and LSR for complex **2** based device. Reproduced with permission from ref (80). Copyright 2018 Royal Society of Chemistry.

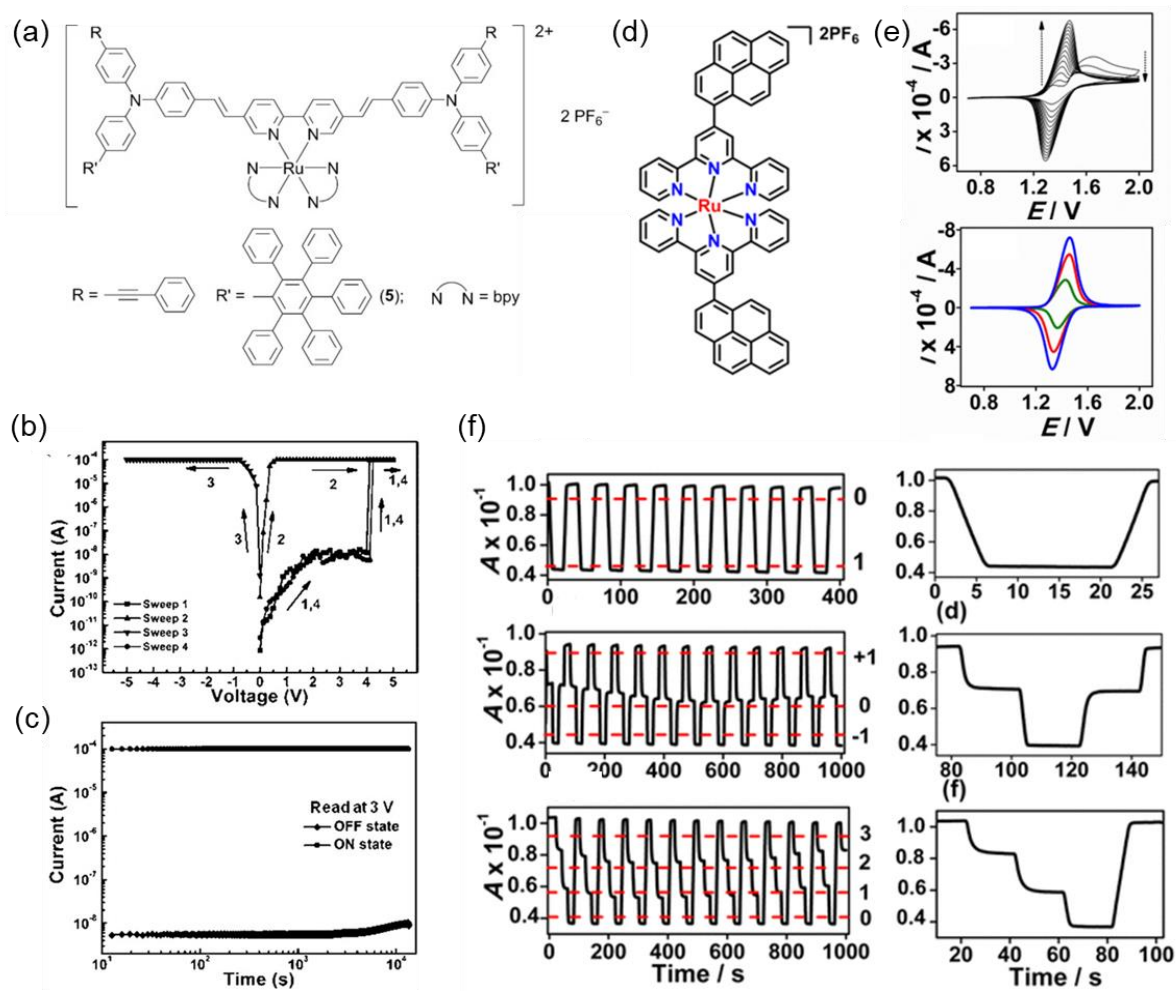

**Figure 20.** (a) Structure of $Ru^{2+}$ complex. (b) Schematic of the memory device based on $Cr^{3+}$ complexes as the active layer. (c) Bipolar switching between HSR and LSR for complex **2** based device. Reproduced with permission from ref (81). Copyright 2016 Wiley Publications. (d) Structure of $Ru^{2+}$ complex. (e) Oxidative electropolymerization of the complex on ITO substrate and CV profiles of the film after 5 (olive) 10 (red) and 20 (blue) electropolymerization cycles. (f) Chrono-absorptiometry switching experiments of $Ru^{2+}$- complex film at λ = 490 nm indicating multiple absorbance states using potential steps of +1.2 V, +1.35 V, +1.45 V and + 1.6 V for realizing binary, ternary and even quaternary memory states. Reproduced with permission from ref (82). Copyright 2016 American Chemical Society.

Kamboj *et al.* reported a square-planar complex $[Co^{2+}L]$ synthesized using the phenalene-derived ligand $LH_2$ = 9,9′-(ethane-1,2-diylbis(azanediyl))bis(1H-phenalen-1-one).[84] Utilizing $[Co^{2+}L]$ as the active material, an ITO/$Co^{2+}$L/Al resistive switching memory device was fabricated and characterized using write-read-erase-read cycles. Impressively, the device exhibited stable and reproducible switching between bistable resistive states for over 2000 seconds highlighting the proposed role of the $Co^{2+}$ metal center and π-conjugated phenalenyl backbone in the redox-resistive switching mechanism.

Contrastingly, Wang *et al.* showed thermoresponsive memory devices, comprising five metallosupramolecular polymers within the PolyCoL1$_x$L2$_y$-PF$_6$ framework (L1 = bisterpyridine; L2 = triterpyridine), were fabricated as Al/[PolyCoL1$_x$L2$_y$-PF$_6$]/ITO devices (**Figure 21a**).[85] These polymers exhibited nonvolatile ternary memory behavior, as evident from the hysteretic *I-V* curves, showcasing transitions among HRS, intermediate-resistance state (IRS), and LRS upon bias voltage addition (**Figure 21b**). Notably, the Al/[PolyCoL1$_{50\%}$L2$_{50\%}$-PF$_6$]/ITO device displayed the most stable and reasonable resistance gap among the various devices evaluated. Furthermore, these memory devices demonstrated dual-stimuli responses—electrical and thermal (**Figure 21c**). Heating the device to 393 K triggered a transition from the LRS to HRS, akin to a warning function, while subsequent cooling restored the LRS. This reversible "LRS–HRS–LRS" transition functioned as a "SAFE–WARNING–SAFE" process, memorizing the heating-cooling cycle. Endurance tests, conducted across varying temperatures, displayed stable performance over ten cycles, confirming the devices' reliability and reusability.

Cui et al. employed the electropolymerization of an asymmetric $Ru^{2+}$ complex (**4+**, **Figure 22a**) that exhibited notable electrochromic behavior and reversible redox activity, demonstrating distinct redox processes at +0.32 and +0.68 V vs. Ag/AgCl.[86] These processes corresponded to three distinct redox states, namely $N^{0/+}$ (ligand-centered), $Ru^{2+/3+}$, and doubly-oxidized states. Films in singly- and doubly-oxidized states exhibited strong absorption peaks at near-infrared (NIR) region at 1070 and 700 nm, respectively (**Figure 22**). These polymeric films displayed promising three-stage near-infrared electrochromism, achieving high contrast ratios (ΔT%) of 52% at 1070 nm and 76% at 700 nm (**Figure 22c**). Notably, the film displayed distinct colors at each redox stage, transitioning from purple to brown to sky blue, respectively. Moreover, the electrochromism featured extended retention times (retain the *transmittance* value after switching off the applied potential) across all three stages, for instance, infinite at −0.20 V, 4 hours at +0.55 V, and 30 minutes at +1.05 V, respectively. Leveraging the singly- and doubly-oxidized states of a 10 nm thick film, the researchers constructed a surface-confined set/reset flip-flop memory, integrating two electrochemical inputs and two near-infrared optical outputs (**Figure 22d**).

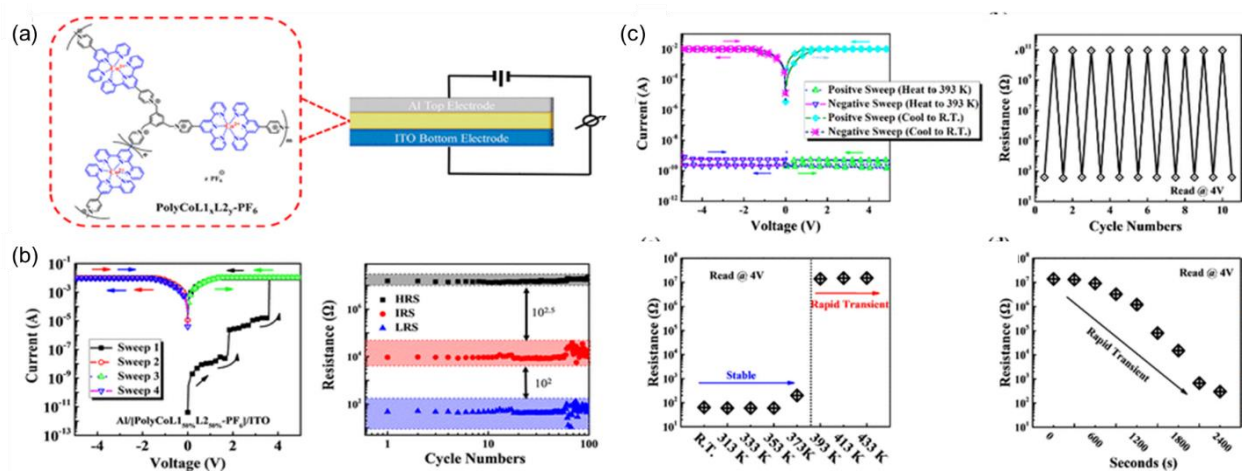

**Figure 21**. (a) Schematic of the memory device using electropolymerized supramolecular assemblies. (b) *I-V* characteristics and endurance properties of Al/[PolyCoL1$_{50\%}$L2$_{50\%}$-PF$_6$]/ITO switching device. (c) Thermoresponsiveness of Al/[PolyCoL1$_{50\%}$L2$_{50\%}$-PF$_6$]/ITO shown by *I-V* characteristics and endurance tests on repetitive cooling and heating. Reproduced with permission from ref (85). Copyright 2017 American Chemical Society.

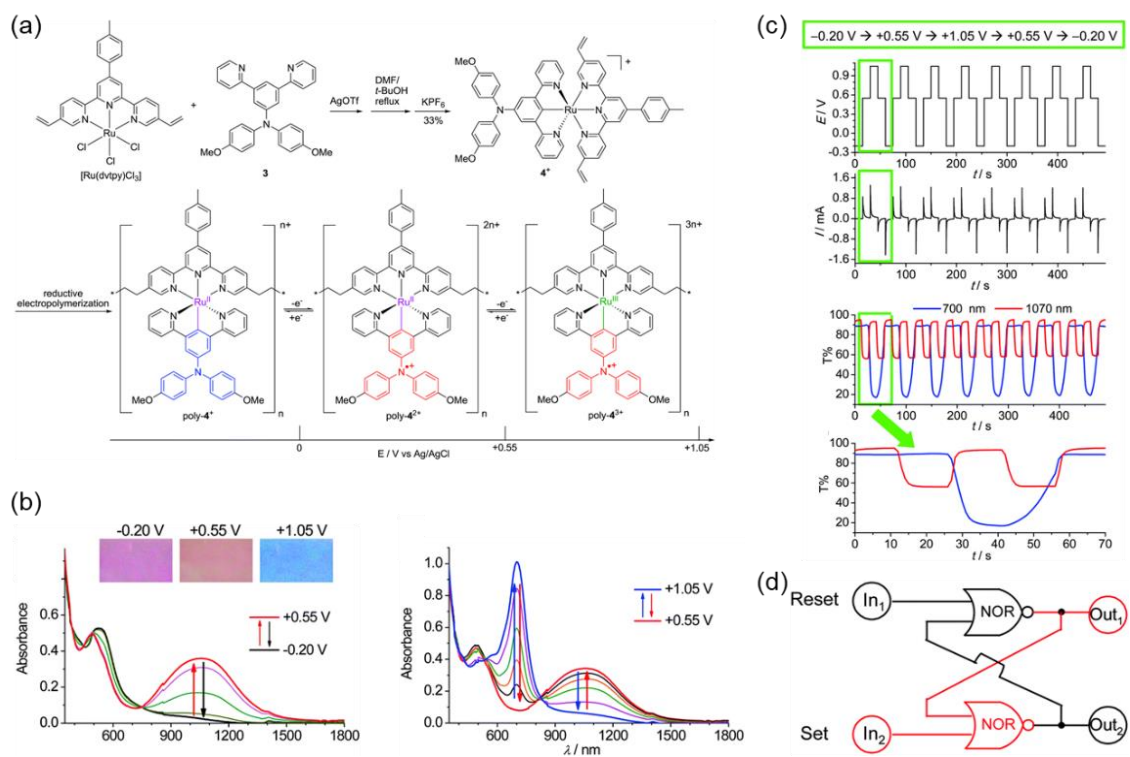

**Figure 22**. (a) Synthesis of asymmetric Ru$^{2+}$ complex (**4$^+$**) for electropolymerization on glassy carbon and ITO-coated glass. (b) NIR spectral switching with applied potential. (c) Electrochromic switching during the four-step potential cycles. (d) Set/Reset flip-flop circuit. Reproduced with permission from ref (86). Copyright 2014 Royal Society of Chemistry.

## 4. Summary and Perspective

(1) In conclusion, this comprehensive overview highlights the remarkable strides made in the field of molecular electronics over the past 60 years. This review elucidates the evolution from the seminal work on molecular rectifiers in 1974 to the present day, encompassing the exploration of various molecular components and advancements, including graphene-based devices. Emphasizing the crucial role of stable electron transport in ensuring the durability of molecular devices, the narrative extends to express concerns about the limitations of silicon technology at the molecular level. The pursuit of breakthroughs in molecular electronics remains a driving force, fueled by the continuous effort to model molecular devices accurately. Looking forward, ambitious plans and technical innovations, such as the integration of nanopore technology, machine learning, and multimode approaches, are deemed essential to deepen our understanding of fundamental transport mechanisms and pave the way for real-world applications. The ultimate aspiration of building molecular integrated circuits, compatible with the existing silicon semiconductor industry, underscores the need for a revolutionary device architecture. Considering these advancements and challenges, this review conveys a sense of optimism, suggesting that the potential of molecular electronics may finally be realized with a concerted effort in the coming years.

(2) In summary, this comprehensive examination of molecular electronics and transition metal complexes unfolds a narrative that spans over six decades, showcasing a captivating journey marked by conceptual breakthroughs, technical advancements, and scientific achievements. From the seminal paper on molecular rectifiers by Ari Aviram and Mark Ratner in 1974 to the present-day exploration of transition-metal–ligand chromophores, the trajectory of molecular electronics has been a story of persistence and evolution. The review underscores the critical importance of stable electron transport across molecular junctions for the durability of molecular devices. With a focus on molecular components such as transistors, diodes, capacitors, insulators, and wires, the research landscape has witnessed significant strides in recent years. The overview also delves into the developments in graphene and graphene-based molecular devices, revealing the intricate tapestry of materials and technologies contributing to the field of molecular electronics. The discussion extends beyond the confines of molecular electronics, touching upon the broader implications for the semiconductor industry. The looming concern that silicon technology may encounter limitations at the molecular level prompts a call for updated theories in molecular electronics. Despite the challenges, the prospect of achieving stable and efficient electron transport in molecular devices propels ongoing efforts. The narrative then shifts to the historical context, acknowledging the intriguing and intriguing nature of d-transition metal chemistry and the growing exploration of f-block complexes. The study of these compounds, driven by the desire to realize molecular electronic and photonic technologies, has given rise to diverse luminescent complexes with potential applications in solar energy conversion, Organic light emitting diodes, and resistive memories.

The paper anticipates that with refined tuning of intermolecular interactions, supramolecular assembly, and controlled patterning, new molecules with unique properties will emerge. This optimism is grounded in the versatility of synthetic methods, the ease of preparation, and the potential for structural optimization in transition metal complexes. Challenges such as lower charge mobilities and less ordered solid-state structures are acknowledged, but the prospect of overcoming these hurdles through innovative approaches remains a driving force. The historical progression concludes by emphasizing the collaborative nature of research, with effective communication between synthetic materials-oriented molecular inorganic chemists, physicists, and device engineers being pivotal for practical applications. While acknowledging that viable commercial devices based on metal complexes are yet to fully materialize, the review expresses confidence that, given the rapid progress witnessed in recent years, the future holds bright prospects for the integration of metal complexes into molecular electronic and photonic devices. As the scientific community continues to navigate these frontiers, the potential for groundbreaking advancements in molecular electronics appears poised for realization.